\documentclass[doublecolumn,manuscript]{aastex}

\usepackage{spr-astr-addons}

\usepackage{mathtools,lipsum,cuted}
\usepackage{multicol}
\usepackage{amsmath}     
\usepackage{amssymb}                       
\usepackage{mathrsfs}                          
\usepackage{upgreek}
\usepackage{bm}  

\shorttitle{Matese-Whitman mass function}

\shortauthors{Shee et al.}

\begin{document}

\title{On the features of Matese-Whitman mass function}

\author{Dibyendu Shee\altaffilmark{1}}
\altaffiltext{1}{Department of Physics,
	Indian Institute of Engineering Science and Technology, Shibpur,
	Howrah 711103, West Bengal, India\\dibyendu\_shee@yahoo.com}

\author{Debabrata Deb\altaffilmark{2}}
\altaffiltext{2}{Department of Physics,
	Indian Institute of Engineering Science and Technology, Shibpur,
	Howrah 711103, West Bengal, India\\d.deb32@gmail.com}

\author{Shounak Ghosh\altaffilmark{3}}
\altaffiltext{3}{Department of Physics,
	Indian Institute of Engineering Science and Technology, Shibpur,
	Howrah 711103, West Bengal, India\\shnkghosh122@gmail.com}

\author{B.K. Guha\altaffilmark{4}}
\altaffiltext{4}{Department of Physics, Indian Institute of
	Engineering Science and Technology, Shibpur, Howrah 711103, West
	Bengal, India\\bkguhaphys@gmail.com}

\author{Saibal Ray\altaffilmark{5}}
\altaffiltext{5}{Department of Physics, Government College of
	Engineering and Ceramic Technology, Kolkata 700010, West Bengal,
	India\\saibal@associates.iucaa.in}

\begin{abstract}
In the present paper we exhaustively examine the physical status of the socalled Matese-Whitman mass function
~[J.J. Matese and P.G. Whitman, Phys. Rev. D, 22, 1270 (1980)]. As a first step, we construct the relevant 
Einstein field equations with an anisotropic matter distribution under the approach of Conformal killing Vector. 
In the intermideate step we find a set of exact solutions by using the Matese-Whitman mass function. 
Eventually we conduct several physical tests to explore features of the applied mass function in connection 
to the specific compact stars. It can be observed that all the features of the model based on the Matese-Whitman mass function are of physical interests.
\end{abstract}


\section{Introduction}
A particular type of mass function, known in the literature as the 
Matese-Whitman mass function~\citep{MW1980}, 
that gives a monotonic decreasing matter density can be provided by
\begin{equation}
m(r)=4\,\pi \int_{0}^{r}\!\rho \left( r \right) {r}^{2}\,{\rm d}r=\frac{br^{3}}{2(1+ar^{2})}, \label{eq0}
\end{equation}
where  $a$ and $b$ are two positive constants.  It is observed that 
this mass function has earlier been used by~\cite{Mak2003} 
to model an anisotropic fluid star,~\citet{Lobo2006} to develop 
a model of dark energy star,~\citet{Sharma2007} to model a 
class of relativistic stars with a linear equation of state, 
and~\citet{Maharaj2009} to strange stars with quark matter. 
Further and specific litarure survey on the Matese-Whitman mass function 
shows that there are some other works done by several 
authors~\citep{Rahaman2010,Bhar2016,Dayanandan2016} 
which are of particular interest. 

The compact anisotropic relativistic astrophysical objects are
always the field of immense interest to the active
astrophysicists. The compact objects are formed  at the end point
of the stellar evolution and its exact nature is still playing
`hide and seek' with the researchers. It is knwon that 
a neutron star-like compact star is the
final stage of a gravitationally collapsed star which, after
exhausting all its thermo-nuclear fuel, gets stabilized by
degenerate pressure.

The first exact solution of Einstein field equations for the
interior of a compact object was obtained by Schwarzschild in
1916, after that several relativists obtained other exact
solutions.~\citet{Delgaty1988} analysed that
out of 127 published solutions only 16 solutions satisfy all
the physical conditions.~\citet{Mak2002} were
obtained an exact solution of Einstein field equations, describing
spherically symmetric and static anisotropic stellar
configuration, by assuming a particular form of anisotropic
factor.~\citet{Ruderman1972} has shown that the nuclear
matter may have anisotropic features at very high density ranges
($> 10^{15}$ gm/c.c.). At this range the nuclear interactions must
be treated relativistically. As a result of the anisotropy,
pressure inside the fluid sphere can be decomposed into two parts
namely radial pressure $p_r$ and tangential pressure $p_t$, where
$p_t$ is in the perpendicular direction to $p_r$. 

It has been argued that anisotropy may
occurs in various reasons, e.g., the existence of external field,
in presence of type $P$ superfluid, rotation, phase transition,
magnetic field, mixture of two fluids, existence of solid core etc.
Including  the effect of local anisotropy,~\citet{Bowers1917} showed that anisotropy may have
non-negligible effects on the parameters like maximum equilibrium
mass and surface redshift.~\citet{Santos1997}
have studied local anisotropy in self gravitating systems. For
modeling the compact astrophysical objects physically and more
realistically (with or without cosmological constant), several
astrophysicists have chosen anisotropic matter distribution which
are in the literature~\citep{Mak2002,Mak2003,Usov2004,Kalam2010,Jafry2010,Maulick2012,Ray2012,Karar2012}. 
According to~\citet{Usov2004} the reason for consideration
of anisotropy within the compact star could be the presence of
strong electric field.~\citet{Egeland2007} studied mass
and radii of neutron stars by incorporating the
existence of cosmological constant proportionality which depends on
the density of vacuum and by using the Fermi equation
of state together with the Tolman-Oppenheimer-Volkov (TOV)
equation.

CKV is a elegant technique by which one can search for the
inheritance symmetry which provides a natural relationship between
geometry and matter through the Einstein field equation. Several
studies have been done on charged or neutral fluid spheres with a
spacetime geometry that admits a conformal symmetry, in the static
as well as non-static cases. Long ago Herrera and his co-workers~
\citep{Herrera1984,Herrera1985a,Herrera1985b,Herrera1985c} 
have extensively studied the interior solutions admitting conformal
motions. However, there are lots of recent works  on the conformal symmetry available in
the following literature~\citep{Ray2008,Rahaman2010,Ghosh2010,
Nandi2011,Bhar2014,Fatima2014,Khadekar2015,Pradhan2015,Das2015,Shee2016}. 

Now under the above historical background our motivation 
and plan of investigation are based on the following steps: in section 2 we have discussed the
Einstein field equations under the non-static conformal symmetry and
their solutions in section 3 for anisotropic matter distribution. In section 4 
by applying the boundary conditions we have found expressions for the constants and the metric potentials. 
Several physical features have been studied in section 5. Remarks on our model are made in section 6.

\section{The Einstein field equations under non-static conformal symmetry}
The highly nonlinear partial differential equations of Einstein's gravity 
can easily be reduced to ordinary differential equations by using 
the above mentioned technique of CKV.

The interior of a star under conformal motion through non-static
Conformal Killing Vector can be represented as in ref.~
\citep{Maartens1990,Coley1994,Lobo2007,Radinschi2010}
\begin{equation}
L_{\xi}g_{ij}=g_{ij;k}\xi^{k}+g_{kj}\xi_{;i}^{k}+g_{ik}\xi_{;j}^{k}=\psi
g_{ij}, \label{eq1}
\end{equation}
where $L$ represents the Lie derivative operator. It gives the
information of the interior gravitational field of a stellar
configuration with respect to the vector field $\xi$ and 
the conformal factor $\psi$. 

Let us consider that our static
spherically symmetric spacetime admits an one parameter group of
conformal motion. A static, spherically symmetric spacetime can be
described by the line element in the standard form as,
\begin{equation}
ds^{2}=-e^{\nu(r)}dt^{2}+e^{\lambda(r)}dr^{2}+r^{2}(d\theta^{2}+sin^{2}\theta
d\phi^{2}), \label{eq2}
\end{equation}
Were $\nu(r)$ and $\lambda(r)$ are the metric potentials and
function of the radial coordinate r only. Here we have considered
$G=1=c$ in geometrized units.

The proposed charged fluid space-time is mapped conformally onto
itself along the direction $\xi$. Here following Herrera and 
his coworkers~\citep{Herrera1984,Herrera1985a,Herrera1985b,Herrera1985c} 
we assume $\xi$ as non-static but $\psi$ to be static as follows:
\begin{equation}
\xi=\alpha(t,r)\partial_{t}+\beta(t,r)\partial_{r}, \label{eq3}
\end{equation}

\begin{equation}
\psi=\psi(r). \label{eq4}
\end{equation}

The above set of Eqs. (1)-(4) give the following expressions for
$\alpha$, $\beta$, $\psi$ and $\nu$ from ref.~
\citep{Maartens1990,Coley1994,Harko2005,Lobo2007,Radinschi2010,Jafry2010}
\begin{equation}
\alpha=A+\frac{1}{2}kt, \label{eq5}
\end{equation}

\begin{equation}
\beta=\frac{1}{2}Bre^{-\frac{\lambda}{2}}, \label{eq6}
\end{equation}

\begin{equation}
\psi=Be^{-\frac{\lambda}{2}}, \label{eq7}
\end{equation}

\begin{equation}
e^{\nu}=C^{2}r^{2}exp\left[-2kB^{-1}\int\frac{e^{\frac{\lambda}{2}}}{r}dr\right],
\label{eq8}
\end{equation}
where $k$, $A$, $B$ and $C$ are arbitrary constants. According to~
\citet{Maartens1990} one can set $A=0$,
$B=1$ and $C=1$ so that by rescaling we can get
\begin{equation}
\alpha=\frac{1}{2}kt, \label{eq9}
\end{equation}

\begin{equation}
\beta=\frac{1}{2}re^{-\frac{\lambda}{2}}, \label{eq10}
\end{equation}

\begin{equation}
\psi=e^{-\frac{\lambda}{2}}, \label{eq11}
\end{equation}

\begin{equation}
e^{\nu}=r^{2}exp\left[-2k\int\frac{e^{\frac{\lambda}{2}}}{r}dr\right].
\label{eq12}
\end{equation}

Without loss of any generality one can choose $A=0$, $B=1$ and
$C=1$. Hence rescaling of $\xi$ and $\psi$ has been done
in the following manner, $\xi \rightarrow B^{-1}\xi$ and $\psi
\rightarrow B^{-1}\psi$ which leaves Eq.~({\ref{eq1}) invariant.

Now, we shall consider the most general energy momentum tensor
compatible with spherically symmetry in the following form as
\begin{equation}
T_{\nu}^{\mu}=(\rho+p_{r})u^{\mu}u_{\nu}+p_{r}g_{\nu}^{\mu}+(p_{t}-p_{r})\eta^{\mu}\eta_{\nu},
\label{eq13}
\end{equation}
with $u^{\mu}u_{\mu}=-\eta^{\mu}\eta_{\mu}=1$ and
$u^{\mu}\eta_{\nu}=0$. Here $u^{\mu}$ is the fluid $4$-velocity
vector and $\eta^{\mu}$ is the space like vector which is
orthogonal to $u^{\mu}$. Also $\rho$ is the matter density, $p_r$
and $p_t$ are respectively the radial and transverse pressure of
the anisotropic fluid distribution. Here $p_t$ is in the
orthogonal direction to $p_r$. Actually the anisotropic fluid
allows the pressure to differ among spatial direction.
$\Delta=p_t-p_r$ is known as anisotropic factor of the spherical
system~\citep{Lake2009,Bhar2015}.

For the metric given in Eq. (2), the Einstein field equations are~
\citep{Ray2008,Rahaman2010,Das2015,Shee2016} given by
\begin{equation}
e^{-\lambda}\left[\frac{\lambda^{\prime}}{r}-\frac{1}{r^{2}}\right]+\frac{1}{r^{2}}=8\pi\rho,
\label{eq14}
\end{equation}

\begin{equation}
e^{-\lambda}\left[\frac{\nu^{\prime}}{r}+\frac{1}{r^{2}}\right]-\frac{1}{r^{2}}=8\pi
p_{r}, \label{eq15}
\end{equation}

\begin{equation}
\frac{1}{2}e^{-\lambda}\left[\frac{1}{2}(\nu^{\prime})^{2}+\nu^{\prime\prime}-\frac{1}{2}\lambda^{\prime}\nu^{\prime}
+\frac{1}{r}(\nu^{\prime}-\lambda^{\prime})\right]=8\pi p_{t}.
\label{eq16}
\end{equation}

Imposing the conformal motion, one can write the stress energy
components in terms of the conformal function as follows
\begin{equation}
8\pi\rho=\frac{1}{r^{2}}(1-\psi^{2}-2r\psi{\psi^{\prime}}),
\label{eq17}
\end{equation}

\begin{equation}
8\pi p_{r}=\frac{1}{r^{2}}(3\psi^{2}-2k\psi-1), \label{eq18}
\end{equation}

\begin{equation}
8\pi p_{t}=\frac{1}{r^{2}}\left[(\psi-k)^{2}+2r\psi{\psi^{\prime}}
\right].\label{eq19}
\end{equation}

\section{Solutions of the field equations}
Now our task is to find out solutions for the above set of the
modified Einstein equations in terms of the conformal motion.
From the mass function~(\ref{eq0}), the matter density can be obtained as
\begin{equation}
\rho={\frac {b \left( a{r}^{2}+3 \right) }{8\pi \left( a{r}^{2}+1 \right) ^{2}}}. \label{eq20}
\end{equation}

\begin{figure*}[thbp]
\centering
\includegraphics[width=0.4\textwidth]{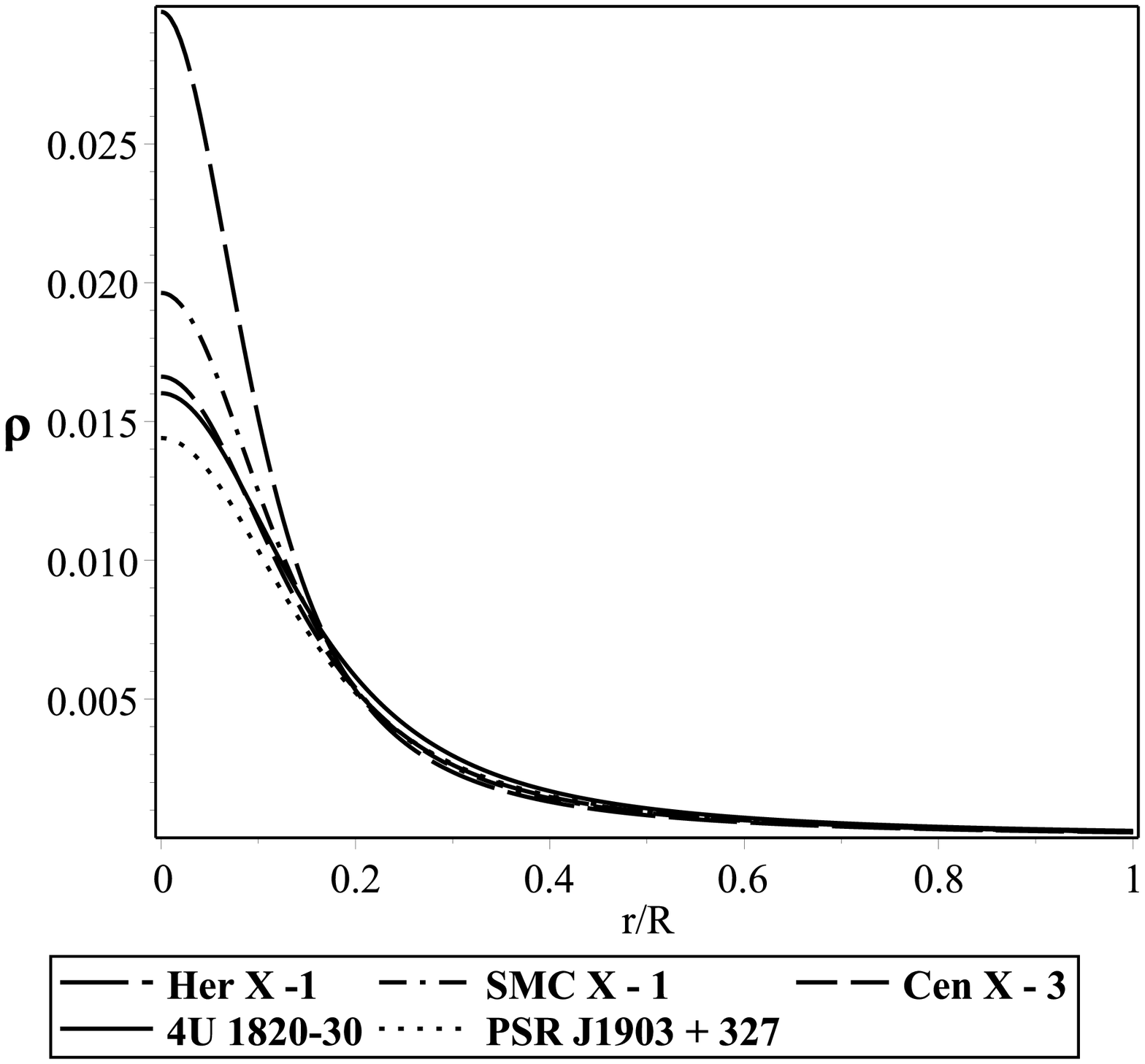}
\includegraphics[width=0.4\textwidth]{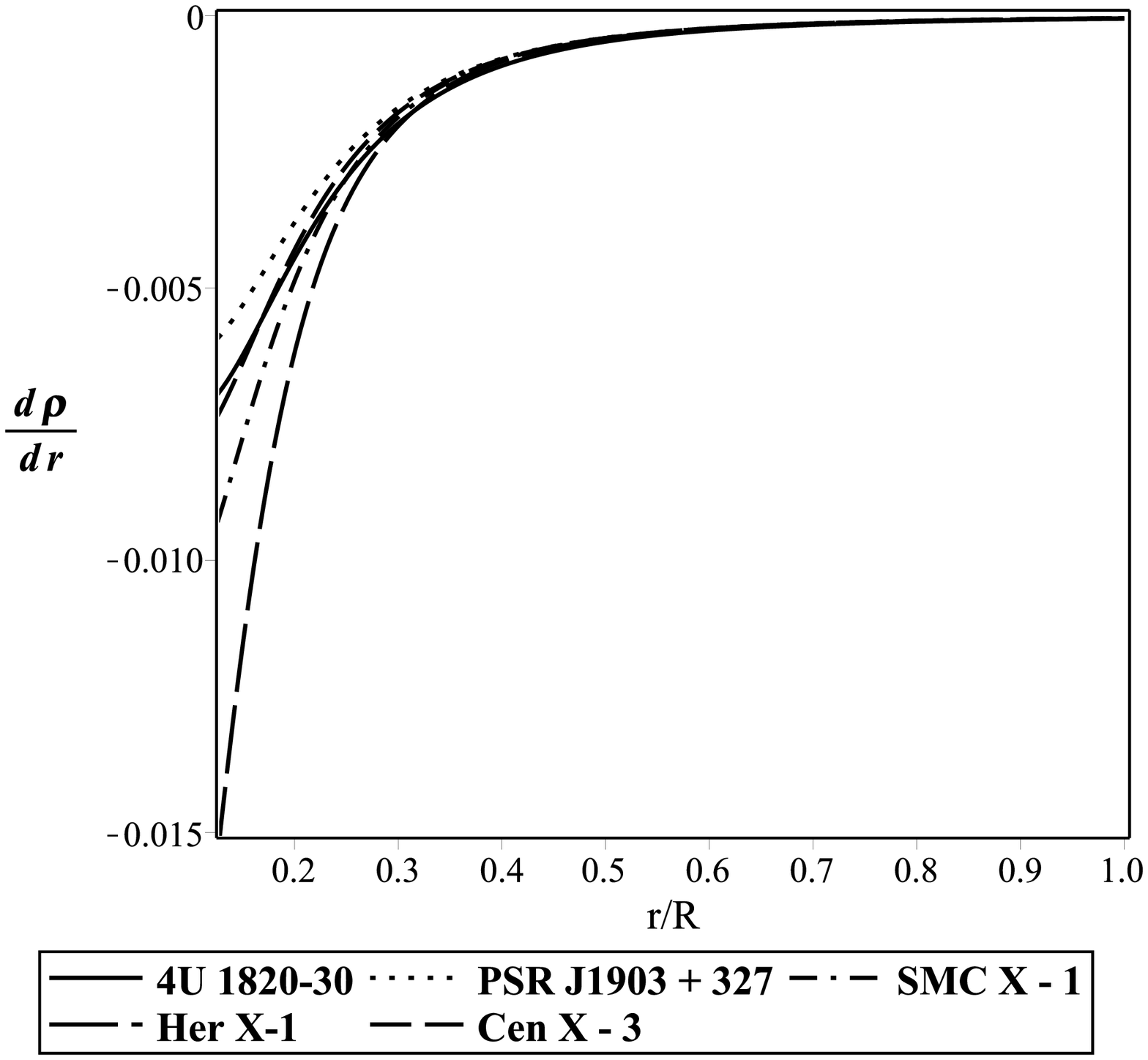}
\caption{Variation of the $\rho$ (left panel) and $\frac{d\rho}{dr}$ (right panel) with the fractional radial coordinate for different compact stars. For plotting this figure we have used the data set of Table 1 which will be later on followed for the other plots also}. \label{den.}
\end{figure*}

The Fig.~\ref{den.} shows that $\rho$ is positive inside the star. It
decreases with the increase of the radius of the star. We also
found from the graph that $\rho {'}<0$ which implies that density
have maximum value at the centre and it decreases monotonically
towards the surface.

Using Eqs. (\ref{eq17}) and (\ref{eq20}) one may have
\begin{equation}
\psi \left( r \right) =\sqrt {1-{\frac {2\,m(r)}{r}}}. \label{eq21}
\end{equation}

Using Eqs. (\ref{eq18})-(\ref{eq21}) we get the expression of $p_r$ and $p_t$ as
\begin{equation}
p_r={\frac {1}{4\,\pi{r}^{2} } \left[ 1-{\frac {3\,b{r}^{2}}{2\,a{r}^{2}
+2}}-k\sqrt {1-{\frac {2\,b{r}^{2}}{2\,a{r}^{2}+2}}} \right] },\label{25}
\end{equation}

\begin{equation}
p_t={\frac {1}{8\pi{r}^{2} } \left[ {k}^{2}-2k\,\sqrt {1-{\frac {2\,b{r
}^{2}}{2\,a{r}^{2}+2}}}-{\frac {b{r}^{2} \left( a{r}^{2}+3 \right) }{
 \left( a{r}^{2}+1 \right) ^{2}}}+1 \right] }.\label{26}
\end{equation}

\begin{figure*}[thbp]
\centering
\includegraphics[width=0.4\textwidth]{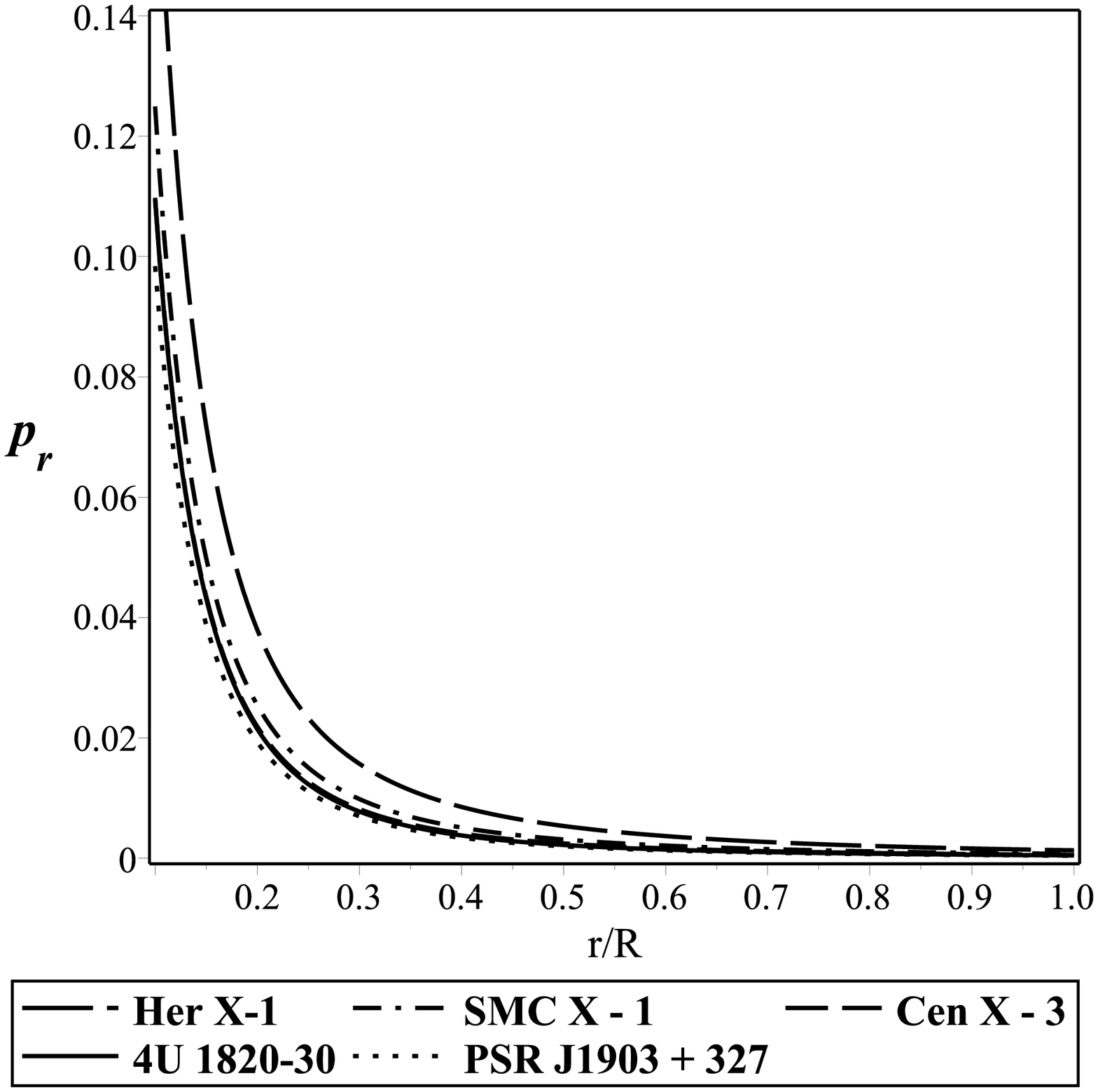}
\includegraphics[width=0.45\textwidth]{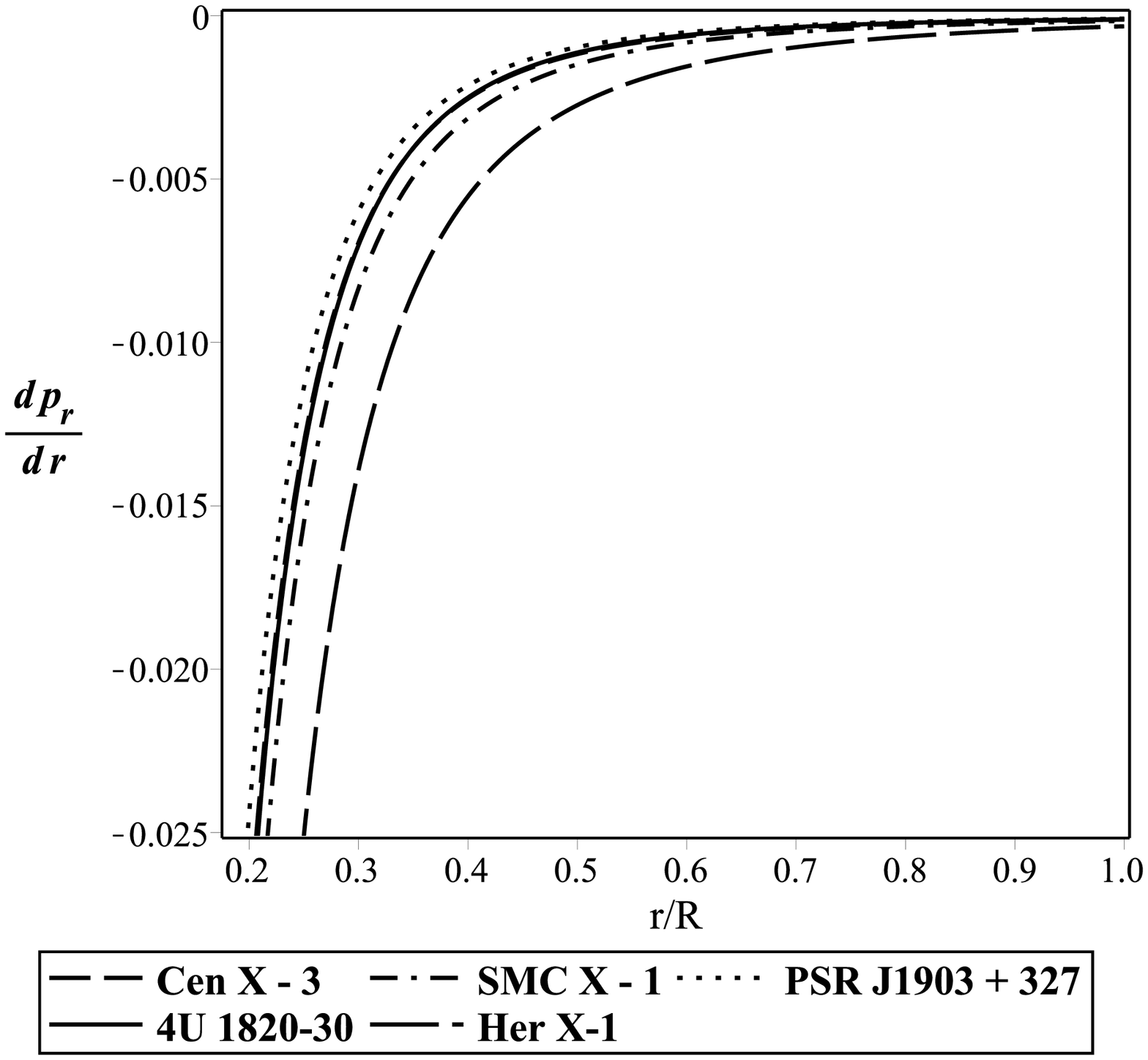}
\caption{Variation of the $p_r$ (left panel) and $\frac{d{p_r}}{dr}$ (right panel) 
with the fractional radial coordinate for different compact stars}\label{pres.}
\end{figure*}

The variation of radial pressure $p_r$ is shown in Fig. \ref{pres.}, which
suffers from a serious problem of singularity. The graph also
shows that $p_r^{'}<0$ which implies radial pressure also
decreases monotonically the radius of the star.

\begin{figure*}[thbp]
\centering
\includegraphics[width=0.4\textwidth]{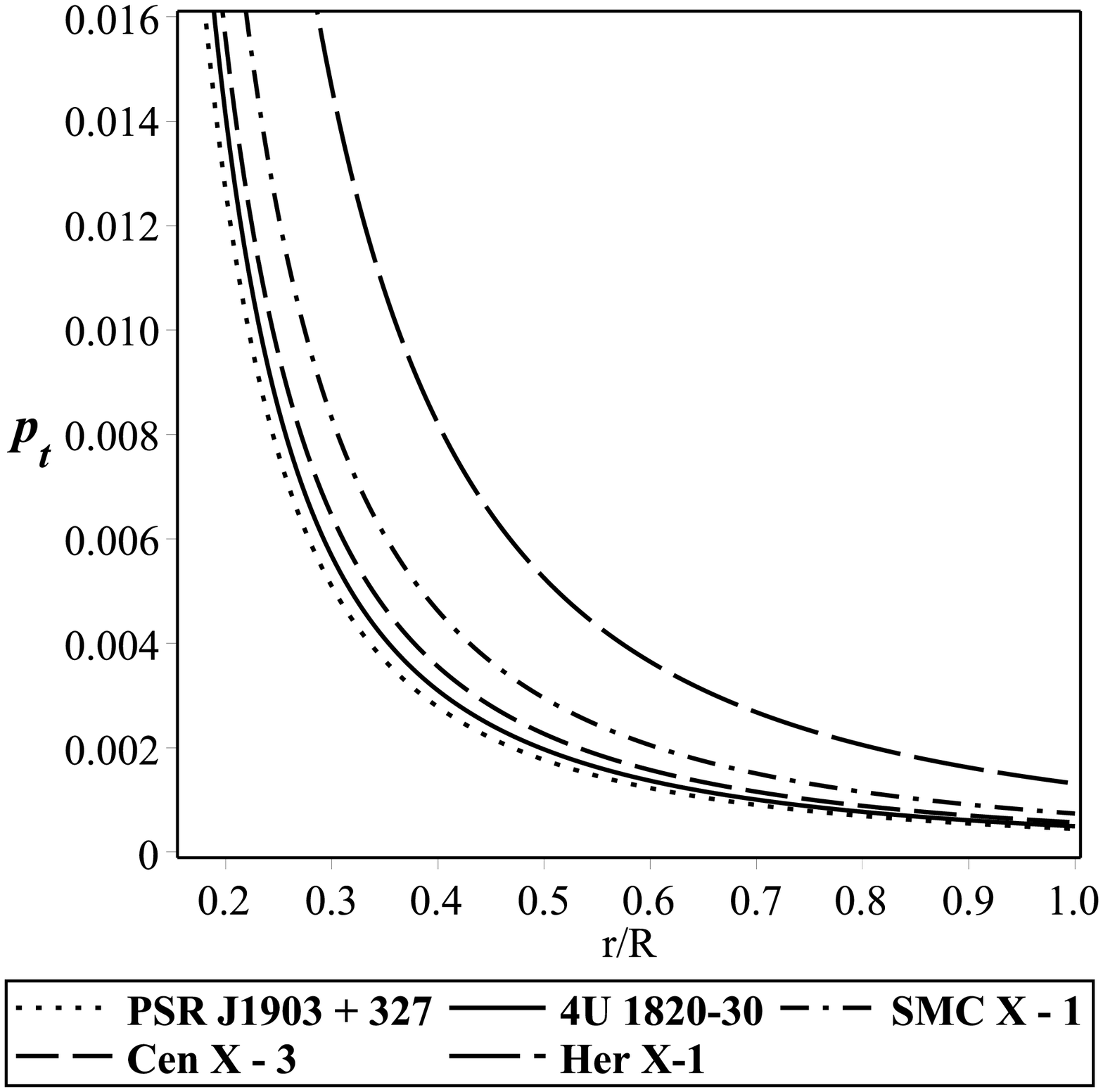}
\includegraphics[width=0.45\textwidth]{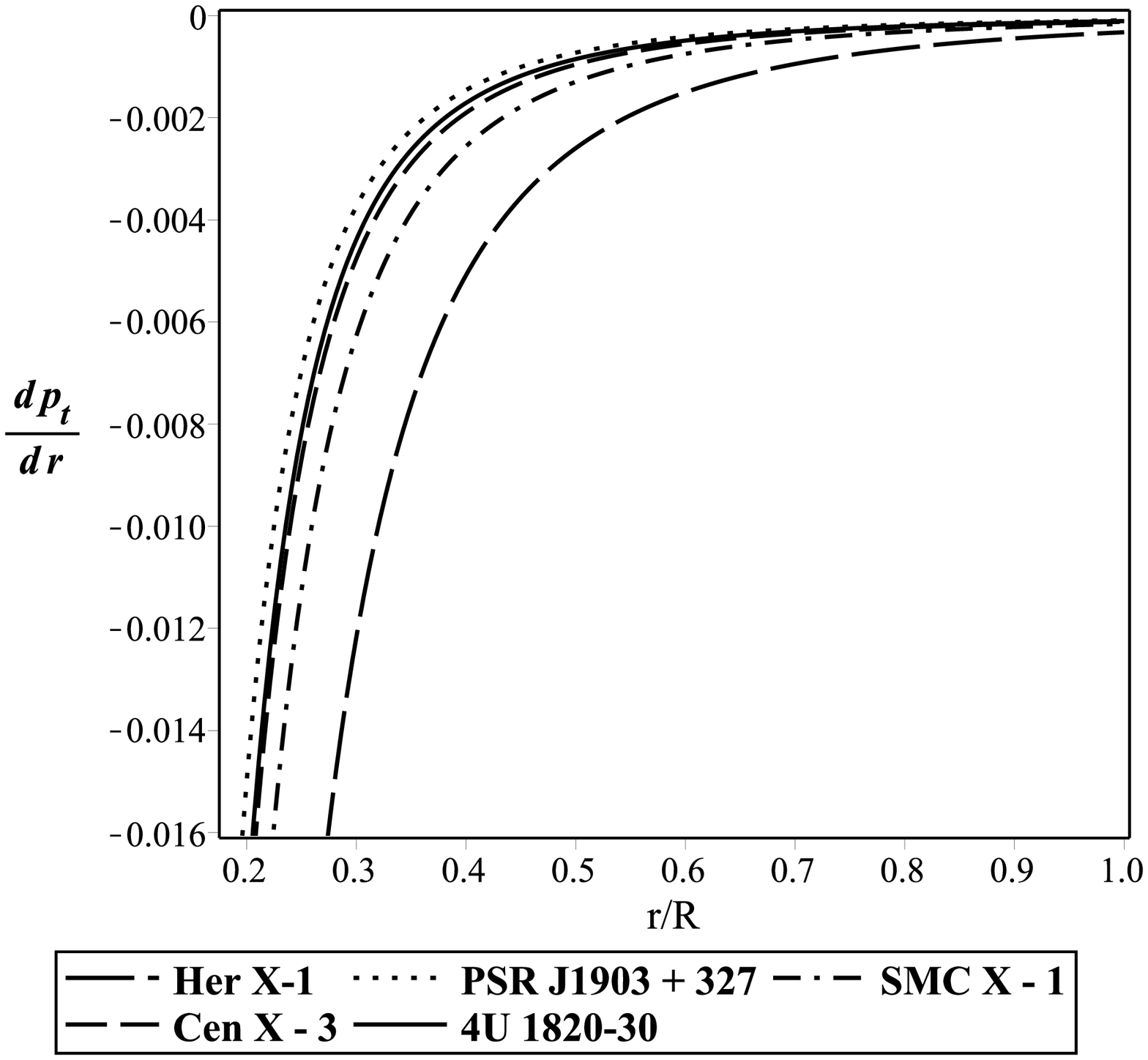}
\caption{Variation of the $p_t$ (left panel), $\frac{d{p_r}}{dr}$ (right panel) 
with the fractional radial coordinatefor different compact stars}\label{tanpr.}
\end{figure*}

The graphical representation of the tangential pressure with the
radius of the star in Fig.~\ref{tanpr.} features that
$p_t$ decreases maintaining the same pattern as $p_r$ and it also
suffers from the singularity problem. $p_t^{'}<0$ implies that it
also monotonically decreasing function of $r$.

The simplest form of the barotropic equation of state (EOS) is
given by $p_i=\omega_i\rho$, where $\omega_i$ are the EOS
parameters along the radial and transverse directions. Actually
this EOS is used for a spatially homogeneous cosmic fluid though
it can be extended to inhomogeneous spherically symmetric
spacetime also. The complicated different forms of $\omega$ with
its possible space and time dependence are available in the
literature~\citep{Zhuravlev2001,Chervon2000,Peebles2003}.

Therefore, the EOS parameter can be written in the following form
as
\begin{equation}
\omega_r=\frac { 2\,\left( a{r}^{2}+1 \right) ^{2}}{b{r}^{2} \left( a{r}^{2}+3
 \right) } \left[ 1-{\frac {3\,b{r}^{2}}{2\,a{r}^{2}+2}}-k\,
\sqrt {1-{\frac {2\,b{r}^{2}}{2\,a{r}^{2}+2}}} \right].\label{27}
\end{equation}

\begin{figure*}[thbp]
\centering
\includegraphics[width=0.4\textwidth]{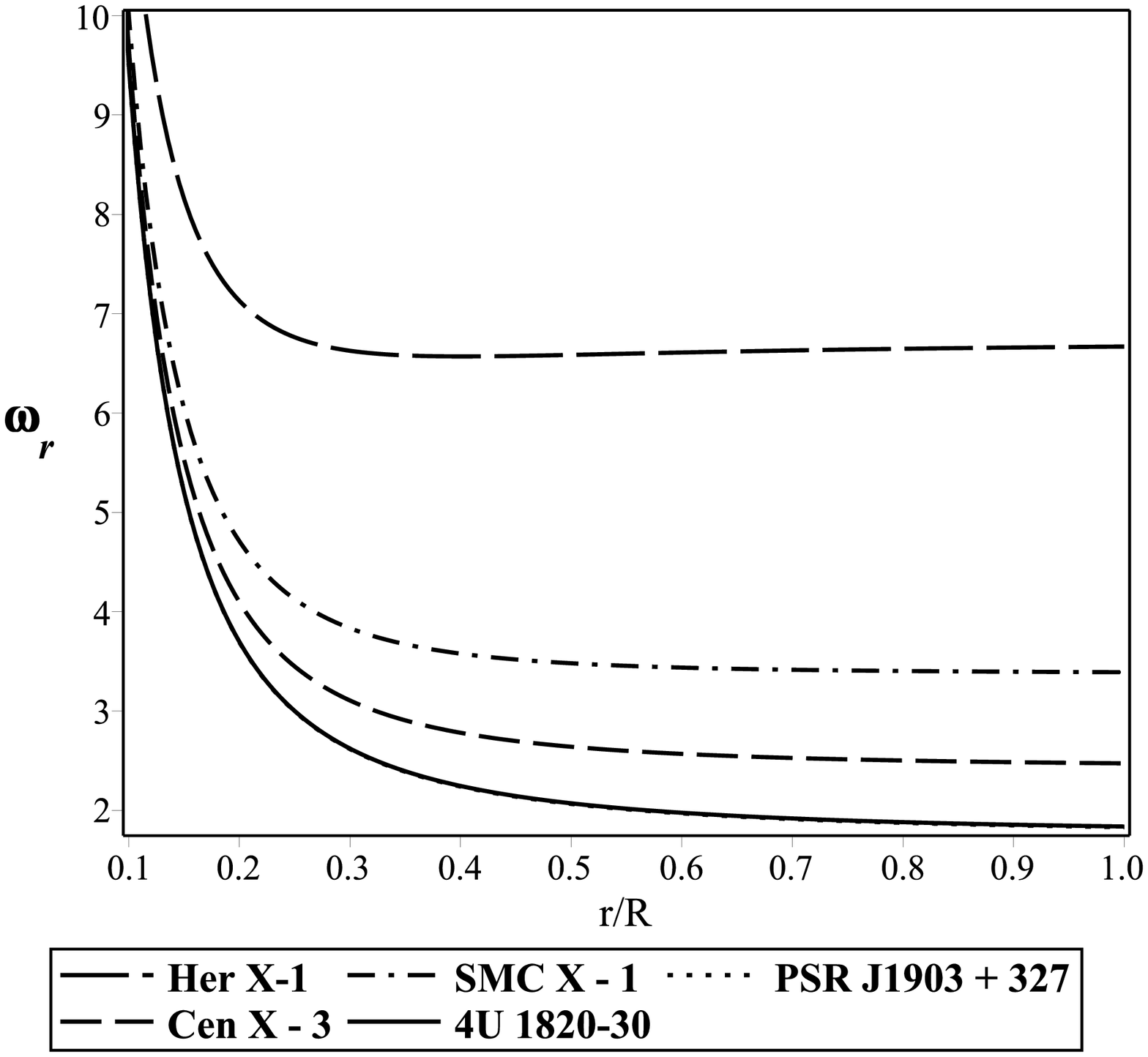}
\includegraphics[width=0.4\textwidth]{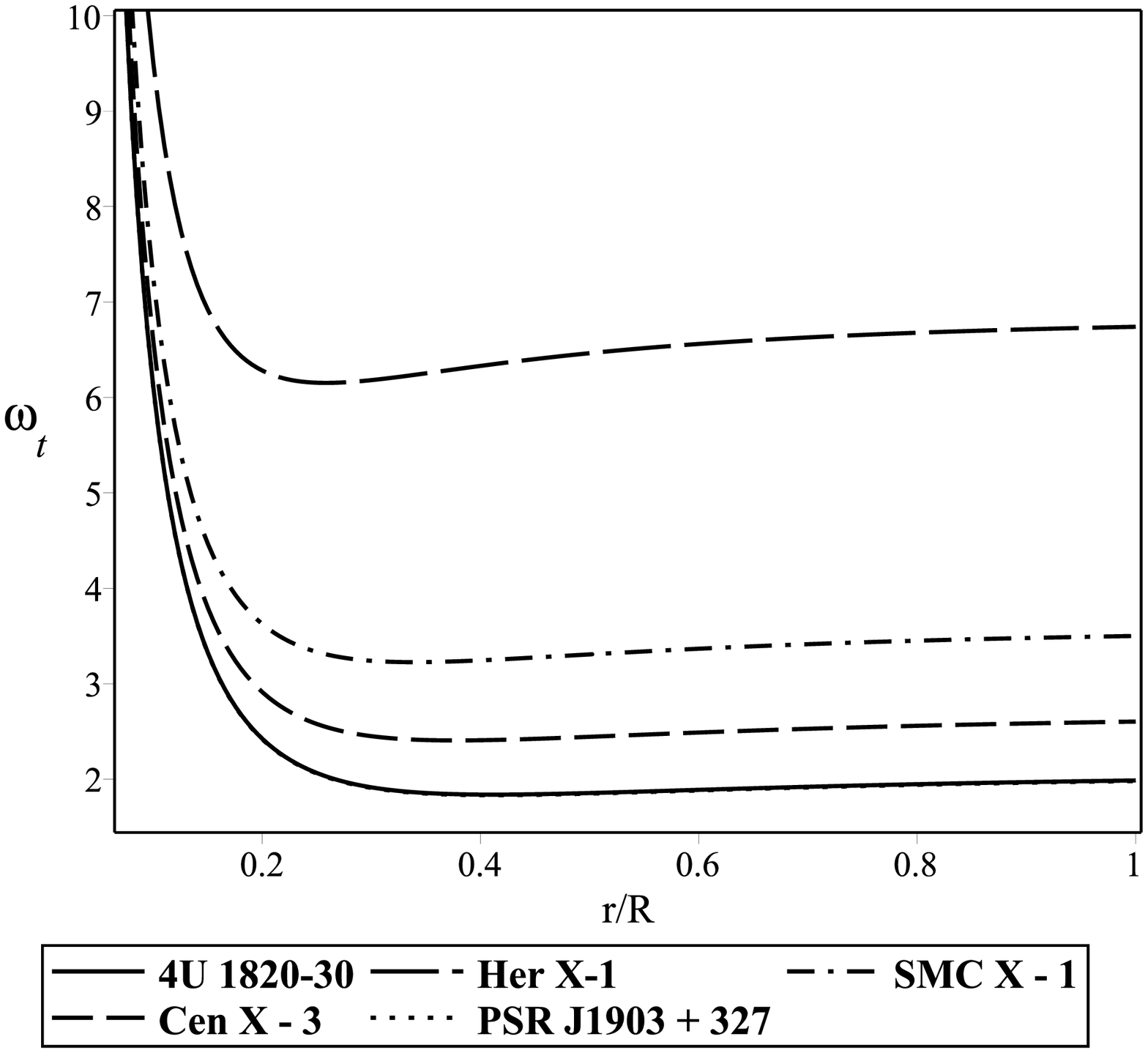}
\caption{Variation of the radial (left panel) and tangential (right panel) 
EOS with the fractional radial coordinate for different compact stars}\label{eos}
\end{figure*}

\begin{eqnarray}
\omega_t=\frac { \left( a{r}^{2}+1 \right) ^{2}}{b{r}^{2} \left( a{r}^{2}+3
 \right)} \times \nonumber \\ \left[k^2-2k\sqrt {1-2{\frac {b{r}^{2}}{2a{r}^{2}
+2}}}  -  {\frac {b{r}^{2} \left(a{r}^{2}+3 \right)}{ \left(a{r}^{2}+1
 \right) ^{2}}}+1 \right].\label{28}
\end{eqnarray}

Fig.~\ref{eos} shows the variation of radial and tangential EOS parameter
with respect to the radial distance. From this plot it is clear
that EOS does not satisfy the condition
$0\leq\omega_r,\omega_t\leq1$. So the underlying matter
distribution is exotic in nature.

\section{Boundary Condition}
Now to derive values of the constants we are matching out interior spacetime with the exterior Schwarzschild metric given as
\begin{eqnarray}
ds^{2} =-\left( 1-\frac{2M}{R}\right) dt^{2}+\left( 1-\frac{2M}{R}\right)^{-1} dr^{2} \nonumber \\+ r^{2} (d\theta ^{2} +\sin ^{2} \theta d\phi ^{2}),\label{BC1}
\end{eqnarray}
where $M$ and $R$ are the effective mass and radius of the stellar system.

Now for a physical stellar the radial pressure should vanish at the surface, 
i.e., at $r=R$ one have ${p_r}=0$. Imposing this condition we get
\begin{equation}
k={\frac {1+ \left( a-3/2\,b \right) {R}^{2}}{\sqrt {a{R}^{2}+1}\sqrt 
{1+ \left( a-b \right) {R}^{2}}}}.\label{BC2}
\end{equation}

Now following the boundary condition for maximized anisotropy 
at the surface of a compact star~\citep{Deb2016a} we have
\begin{equation}
 b=-{\frac {\left( {k}^{2}-1
 \right) \left(a {R}^{2}+1 \right) ^{3}  }{2a{R}^{4} \left(a {R}^{2}-1 \right) }}.\label{BC3}
\end{equation}

Again continuity of the metric potential $e^{\lambda}$ at the surface, $r=R$, one can obtain
\begin{equation}
{\frac {M}{R}}={\frac {b{R}^{2}}{2a{R}^{2}+2}}.\label{BC4}
\end{equation}

Hence from Eqs.~(\ref{BC3}) and (\ref{BC4}) we have several values of $a$ 
and we are taking only that value of $a$ 
for which we are getting physically acceptable solution given as 
\begin{equation}
a={\frac {-R{k}^{2}+2\,M+2\,\sqrt {-2\,MR{k}^{2}+{M}^{2}+2\,MR}+R}{{R}
^{2} \left( R{k}^{2}+4\,M-R \right) }}.\label{BC5}
\end{equation}

\begin{figure*}[thbp]
\centering
\includegraphics[width=0.4\textwidth]{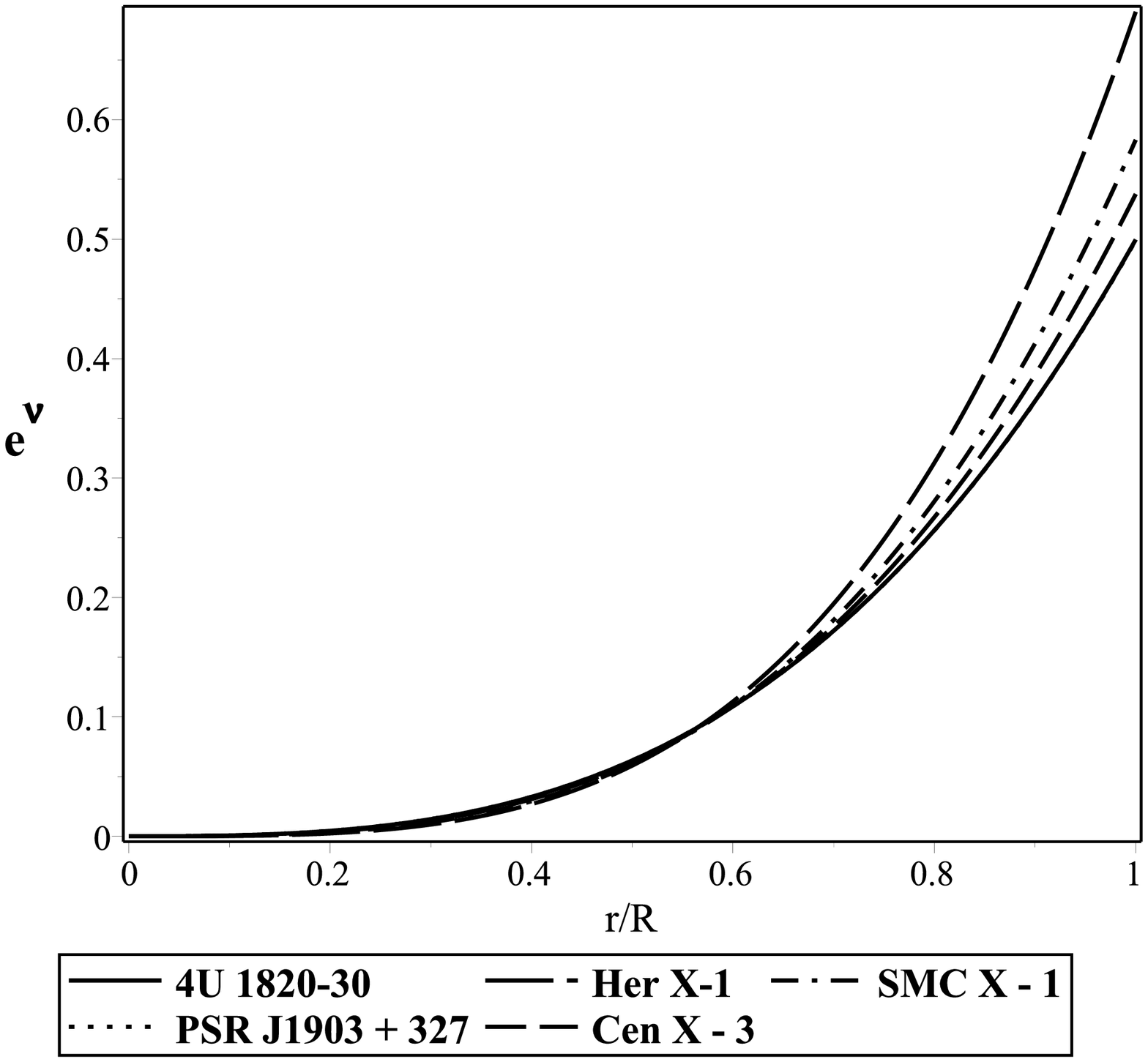}
\includegraphics[width=0.4\textwidth]{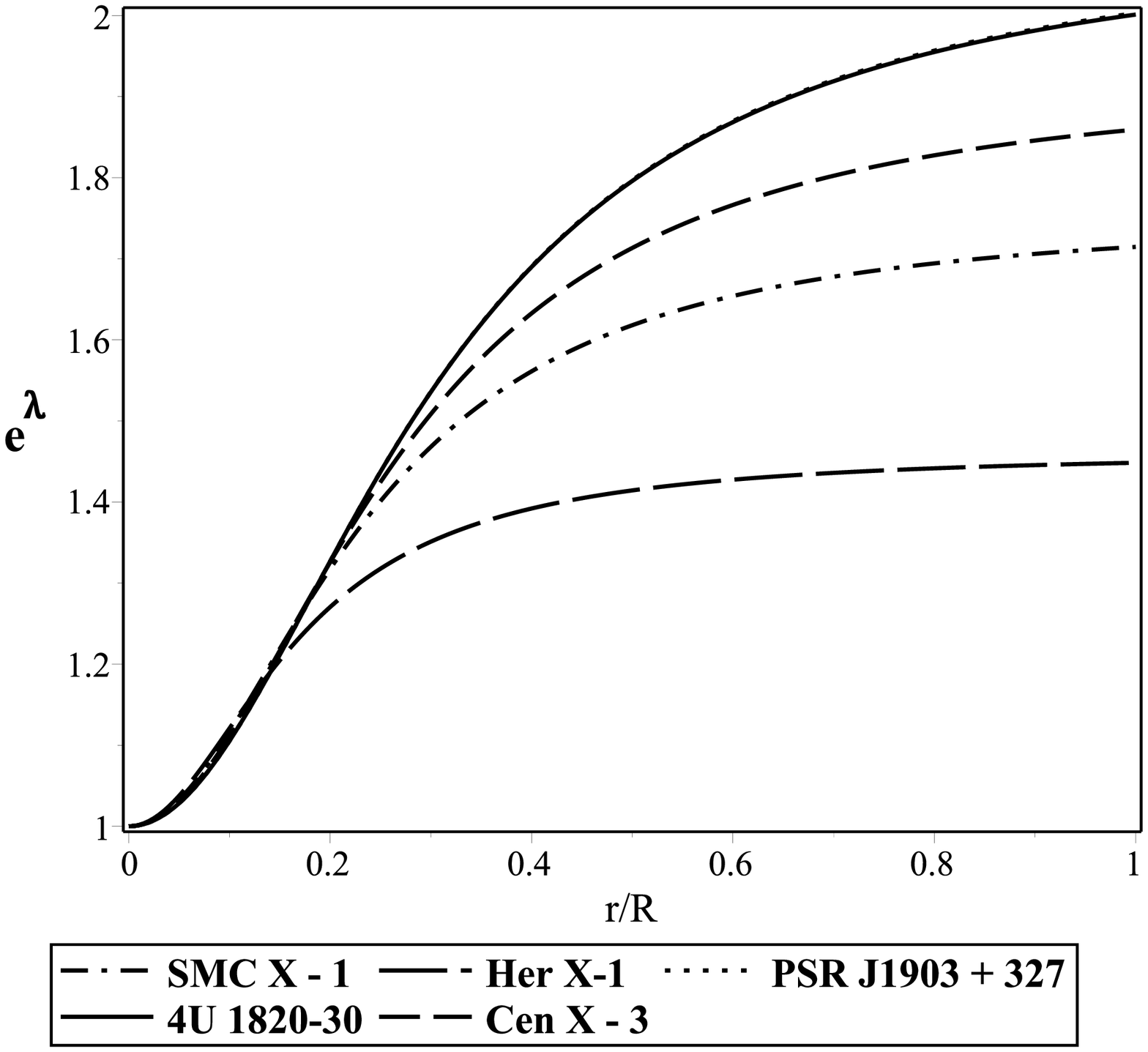}
\caption{Variation of the metric potentials $e^{\nu(r)}$ (left panel) and $e^{\lambda(r)}$ (right panel) 
with the fractional radial coordinate for different compact stars}\label{potentials}
\end{figure*}

The metric potential $e^{\lambda}$ can in a straight forward way be derived using Eqs. (\ref{eq11}), (\ref{eq20}) and (\ref{eq21}) as
\begin{equation}
e^{\lambda}={\frac {a{r}^{2}+1}{1+ \left( a-b \right) {r}^{2}}}.\label{28a}
\end{equation}

In a similar way, the metric potential $e^{\nu}$ can after some manipulation be derived using Eqs. (\ref{eq12}), (\ref{28a}) and (\ref{BC1}) as
\begin{eqnarray}
e^{\nu}= exp\Bigg[{\frac {k\sqrt {a}}{\sqrt {a-b}}\ln~D} - k~{\rm arctanh}~E + k~{\rm arctanh}~F \nonumber \\
+{\frac {1}{\sqrt {a}\sqrt {a-b}}\ln  \left\lbrace {\frac { \left( R-2M \right) {r}^{2}}{{R}^{3}}} \right\rbrace }\Bigg],
\end{eqnarray}
where\\ $D= \left( {\frac {2{a}^{2}{R}^{2}-
2ab{R}^{2}+2\sqrt {1+ \left( a-b \right) {R}^{2}}\sqrt {{R}^{2}a+1
}\sqrt {a-b}\sqrt {a}+2a-b}{2{a}^{2}{r}^{2}-2\,ab{r}^{2}+2\sqrt 
{1+ \left( a-b \right) {r}^{2}}\sqrt {a{r}^{2}+1}\sqrt {a-b}\sqrt {a}+
2a-b}} \right)$, $E= \left({\frac {1+ \left( a-b/2
 \right) {R}^{2}}{\sqrt {1+ \left( a-b \right) {R}^{2}}\sqrt {{R}^{2}a
+1}}}\right)$ and $F=\left({\frac {1+ \left( a-b/2 \right) {r}^{
2}}{\sqrt {1+ \left( a-b \right) {r}^{2}}\sqrt {a{r}^{2}+1}}}\right)$.

Features of variation of the metric potentials are shown in Fig. 5.

\section{Physical features of the model}

\subsection{Anisotropic behaviour}
The measure of anisotropy in pressure can be obtained as
\begin{eqnarray}
\Delta = (p_t-p_r)=\frac {1}{4\pi{r}^{4} } \times \nonumber \\ \left[ {\frac{b{r}^{3} \left( a{r}^{2}-3
 \right) }{ \left( a{r}^{2}+1 \right) ^{3}}} - k^2 r + {\frac{5b{r
}^{3} \left( a{r}^{2}+3 \right) }{2\left( a{r}^{2}+1 \right) ^{2}}}-{\frac {9
\,b{r}^{3}}{2a{r}^{2}+2}}+r \right]. \label{29}
\end{eqnarray}

It can be seen that the `anisotropy' will be directed  outward
when $p_t>p_r$ i.e. $\Delta>0$, and inward when $p_t<p_r$ i.e.
$\Delta<0$. Fig.~\ref{anis.} shows the variation of anisotropy with the radius of the
star. From the figure it is clear that in general $\Delta>0$, i.e., $p_t>p_r$ 
though there is a change over at $r=6.50$~km.
This result implies that anisotropic force is repulsive in nature
and according to~\cite{Gokhroo1994} it helps to construct more compact object. 
It is also to be noted that the anisotropic force dose not vanishes at the centre 
as expected to develop the model of a star.

\begin{figure*}[thbp]
\centering
\includegraphics[width=0.4\textwidth]{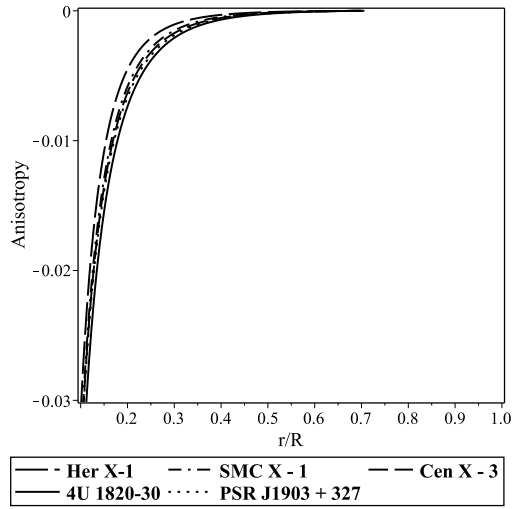}
\caption{Variation of the anisotropic behaviour with the fractional radial coordinate 
for different compact stars}\label{anis.}
\end{figure*}

\subsection {Energy condition}
For an anisotropic fluid sphere the energy conditions namely Weak
Energy Condition (WEC), Null energy Condition (NEC), Strong Energy
Condition (SEC) and Dominant Energy Condition (DEC) are satisfied
if and only if the following inequalities hold simultaneously by
every points inside the fluid sphere.
\begin{equation}
NEC: \rho+p_r \geq 0,
\end{equation}\label{eqn30}
\begin{equation}
 WEC_r: \rho+p_r \geq 0,  \rho> 0,
\end{equation}\label{eqn31}
\begin{equation}
 WEC_t: \rho+p_t \geq 0,  \rho> 0,
\end{equation}\label{eqn31}
\begin{equation}
 SEC: \rho+p_r \geq 0,  \rho+p_r+2p_t> 0,
\end{equation}\label{eqn32}

We shall prove the inequalities with the help of graphical
representation.

\begin{figure*}[thbp]
\centering
\includegraphics[width=0.45\textwidth]{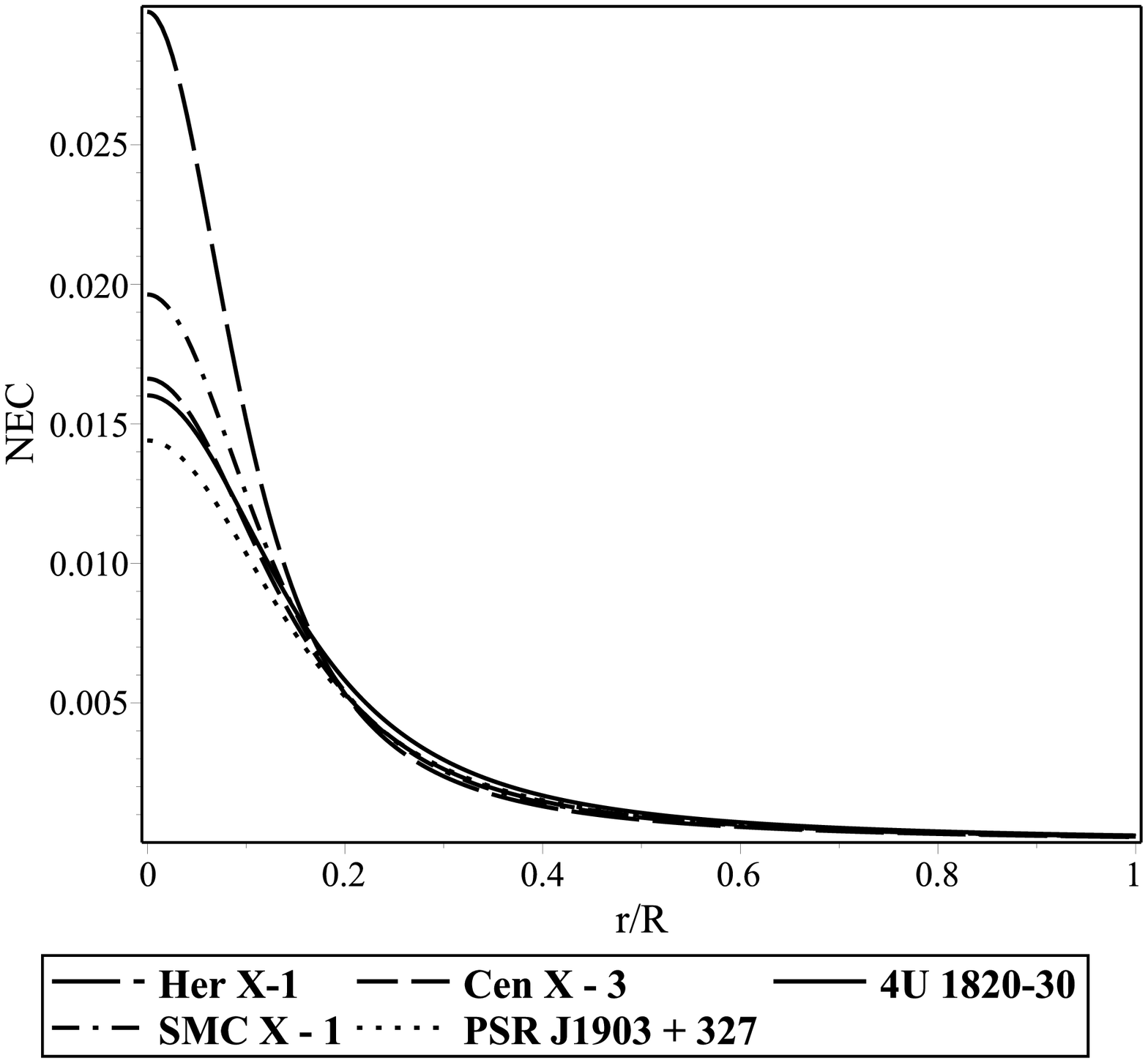}
\includegraphics[width=0.45\textwidth]{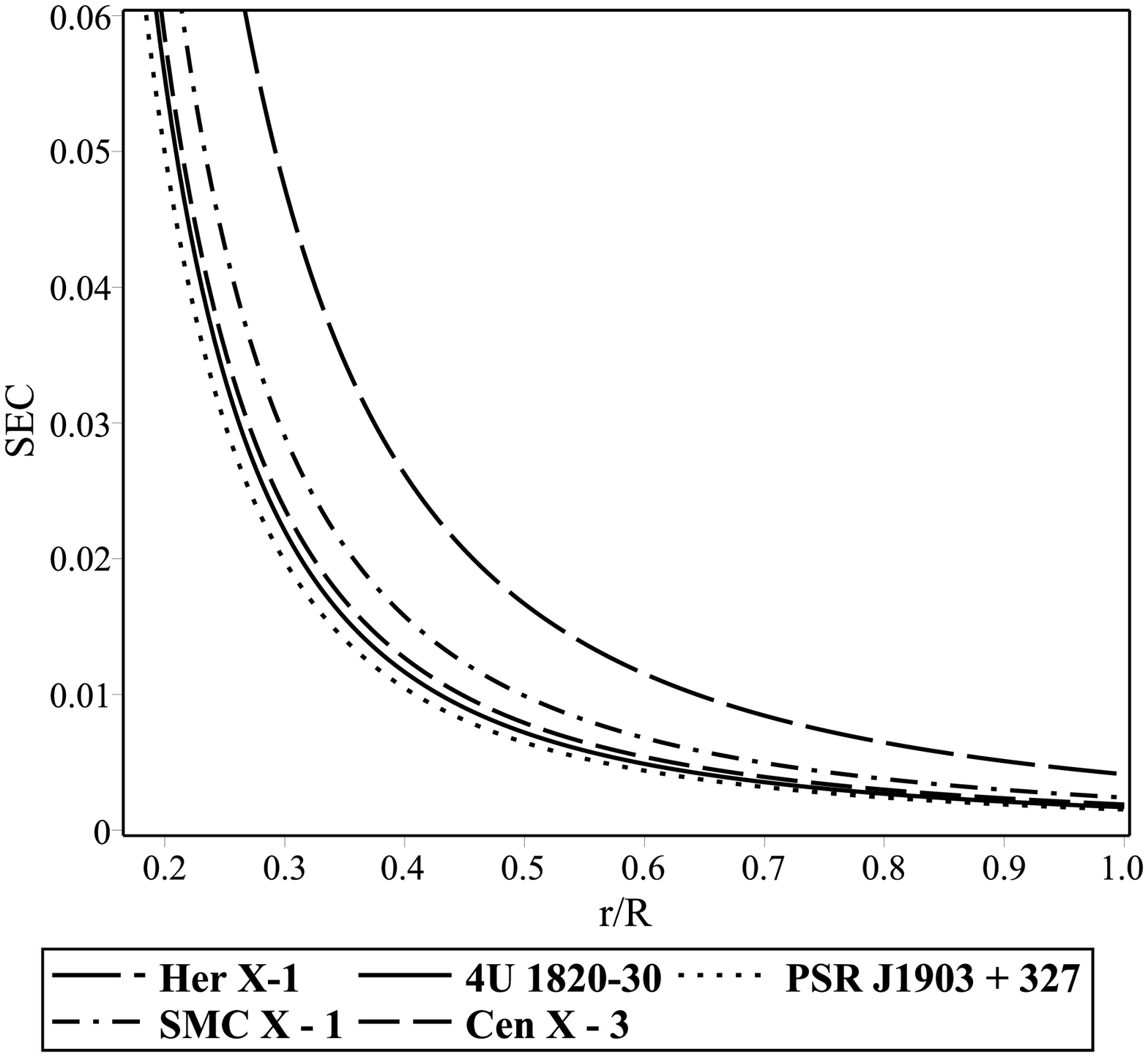}
\includegraphics[width=0.45\textwidth]{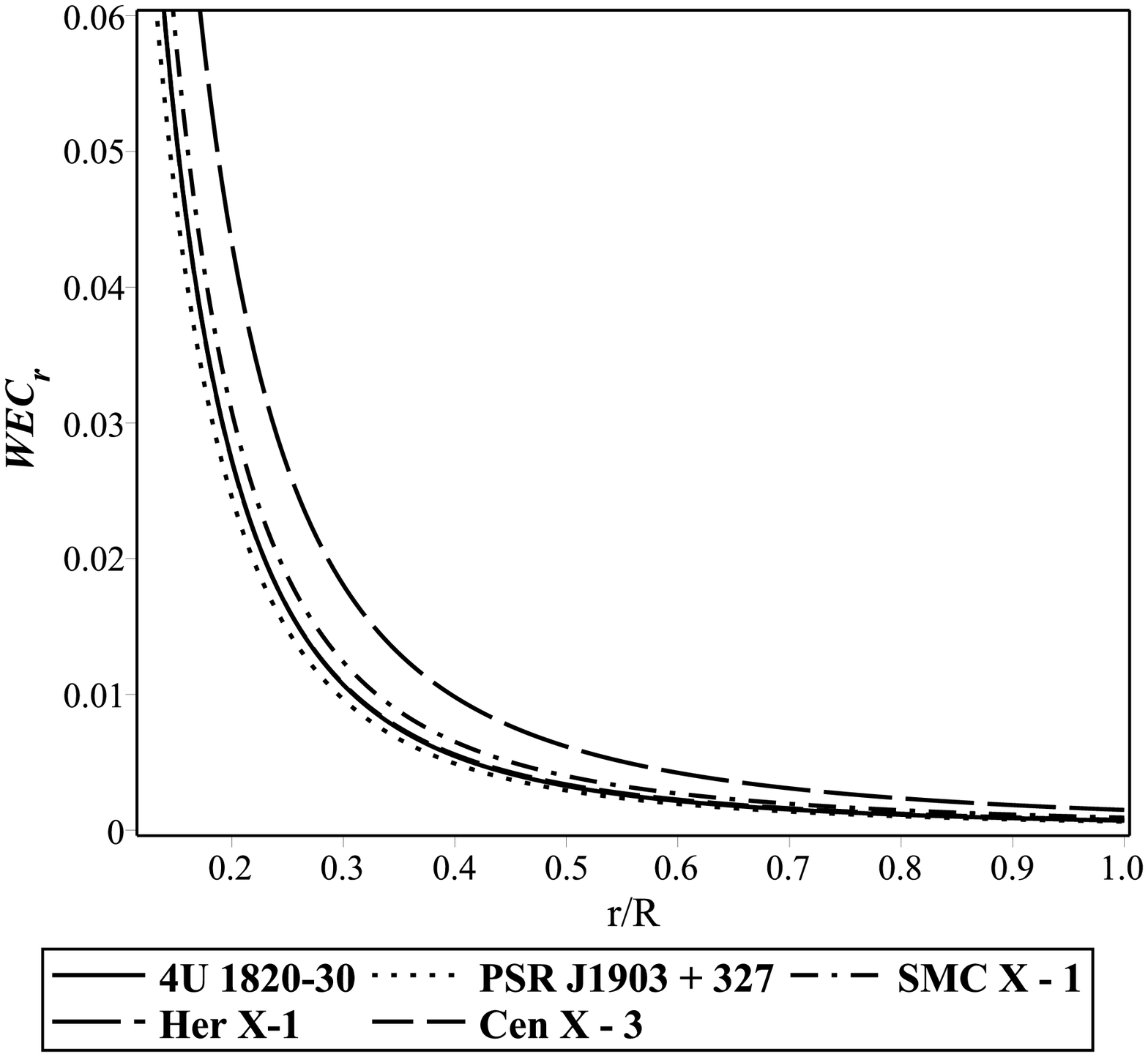}
\includegraphics[width=0.45\textwidth]{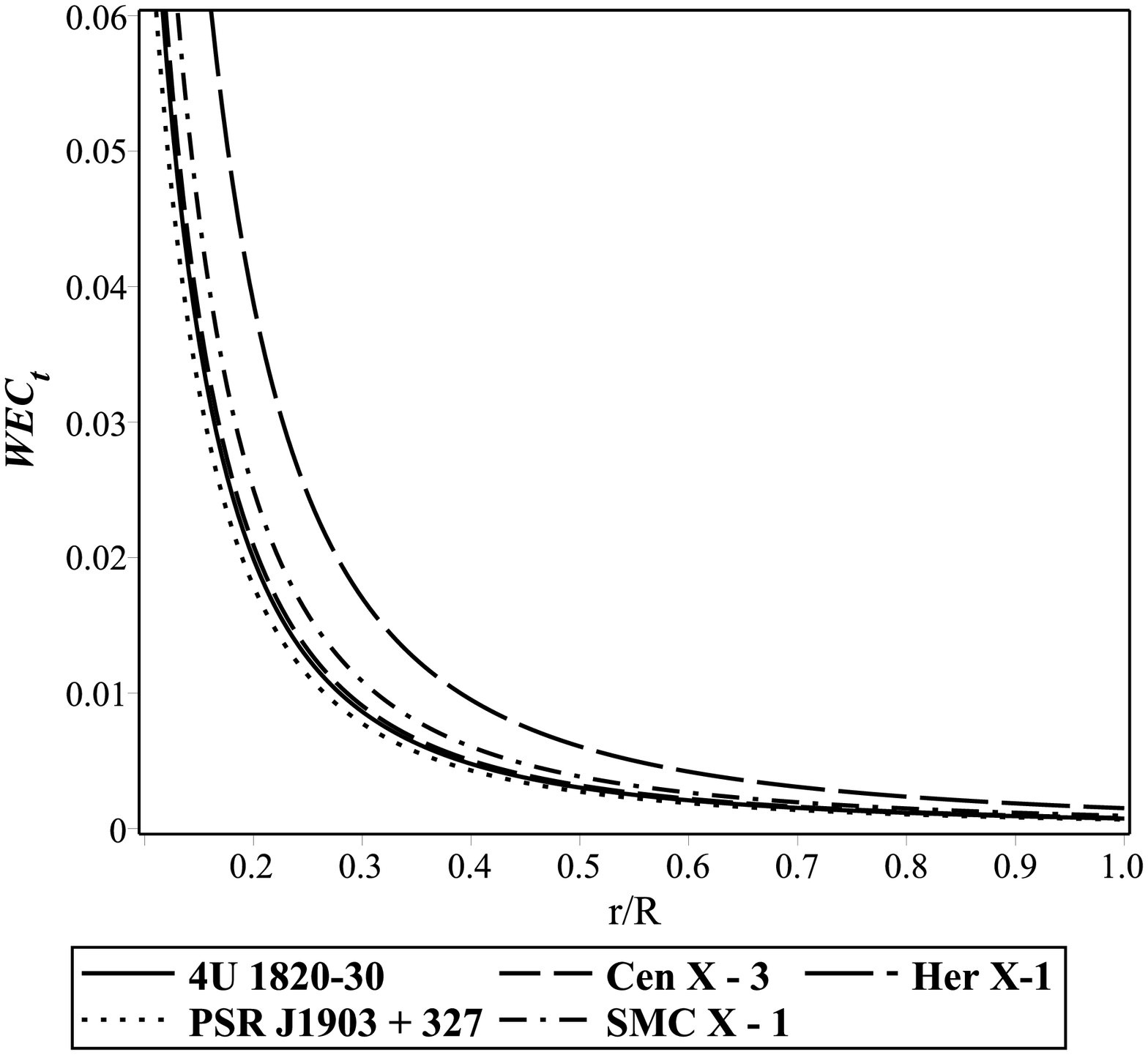}
\caption{Variation of the energy conditions with the fractional radial coordinate 
for different compact stars}\label{energycon.}
\end{figure*}

From Fig.~\ref{energycon.} it is very clear that NEC, WEC, SEC are satisfied by
our model.

\subsection{TOV equation}
To search equilibrium situation of this anisotropic star under
different forces, the generalised Tolman-Oppenheimer-Volkoff (TOV)
equation can written as
\begin{equation}
-\frac{M_G(\rho+p_r)}{r^{2}}e^{\frac{\lambda-\nu}{2}}-\frac{dp_r}{dr}+\frac{2}{r}(p_t-p_r)=0, \label{34}
\end{equation}
where $M_G=M_G(r)$ is the effective gravitational mass inside a
sphere of radius $r$ which can be derived from the Tolman-Whittaker
formula as
\begin{equation}
M_G(r)=\frac{1}{2}r^{2}e^{\frac{\nu-\lambda}{2}}\nu'. \label{35}
\end{equation}

Substituting Eq. (33) in Eq. (32) we get the following form of
the TOV equation
\begin{equation}
-\frac{\nu^{'}(\rho+p_r)}{2}-\frac{dp_r}{dr}+\frac{2}{r}(p_t-p_r)=0.
\end{equation}

The Eq. (34) can be rewritten as
\begin{equation}
F_g+F_h+F_a=0.
\end{equation}
where
\begin{equation}
F_g=-\frac{\nu'}{2}(\rho+p_r),
\end{equation}

\begin{equation}
F_h=-\frac{dp_r}{dr},
\end{equation}

\begin{equation}
F_a=\frac{2}{r}(p_t-p_r).
\end{equation}

\begin{figure*}[thbp]
\centering
\includegraphics[width=0.45\textwidth]{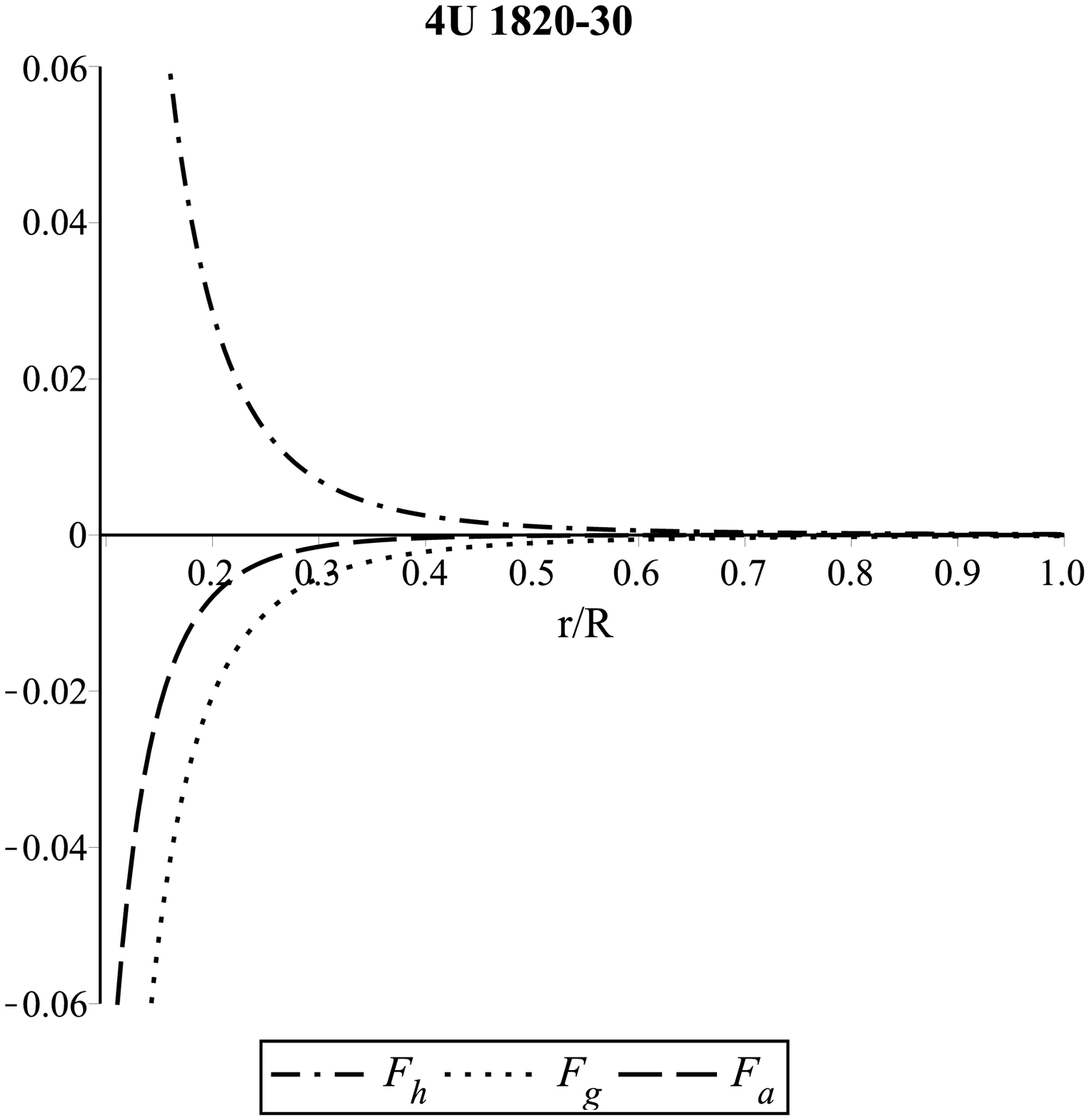}
\includegraphics[width=0.45\textwidth]{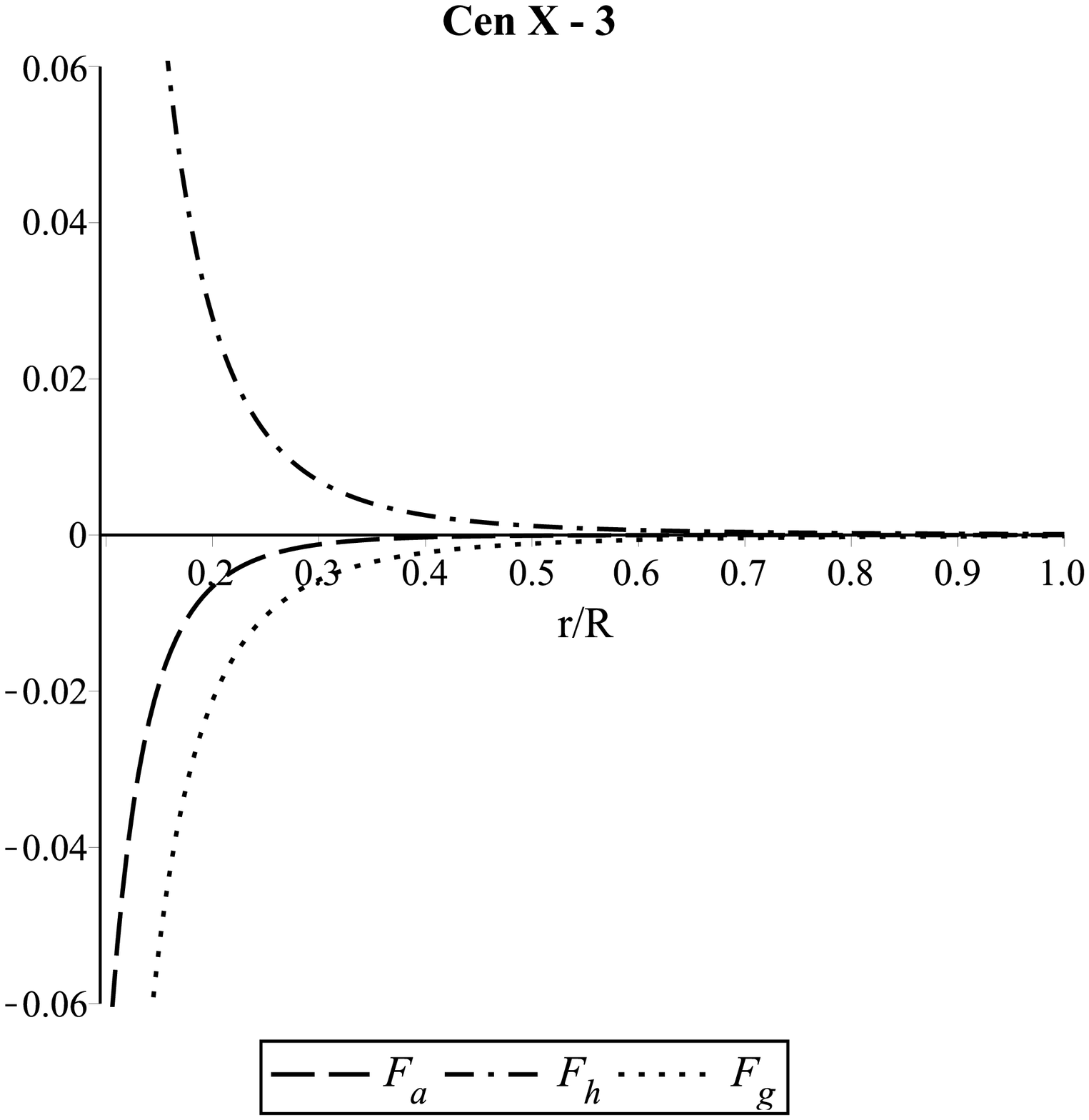}
\includegraphics[width=0.45\textwidth]{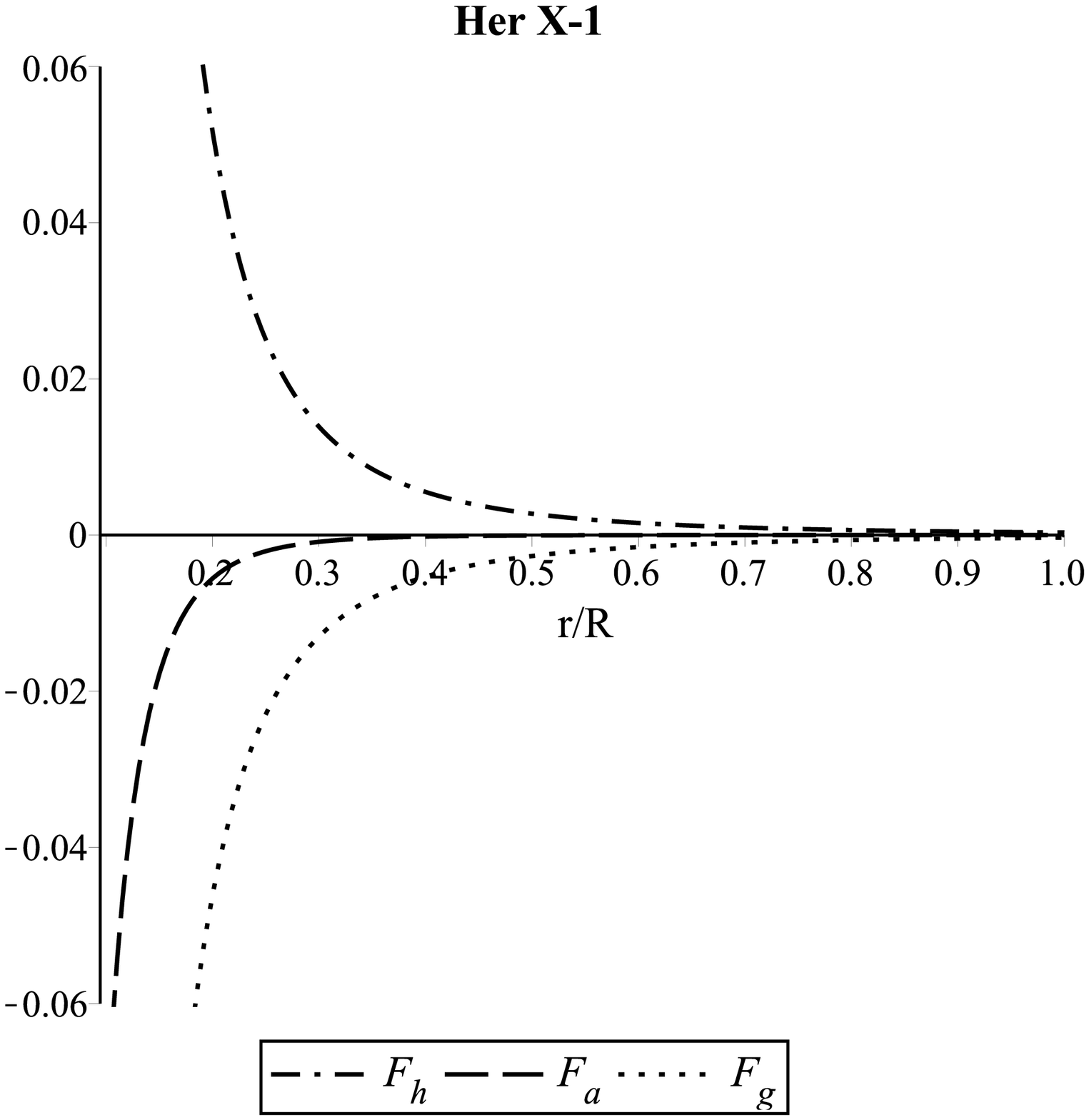}
\includegraphics[width=0.45\textwidth]{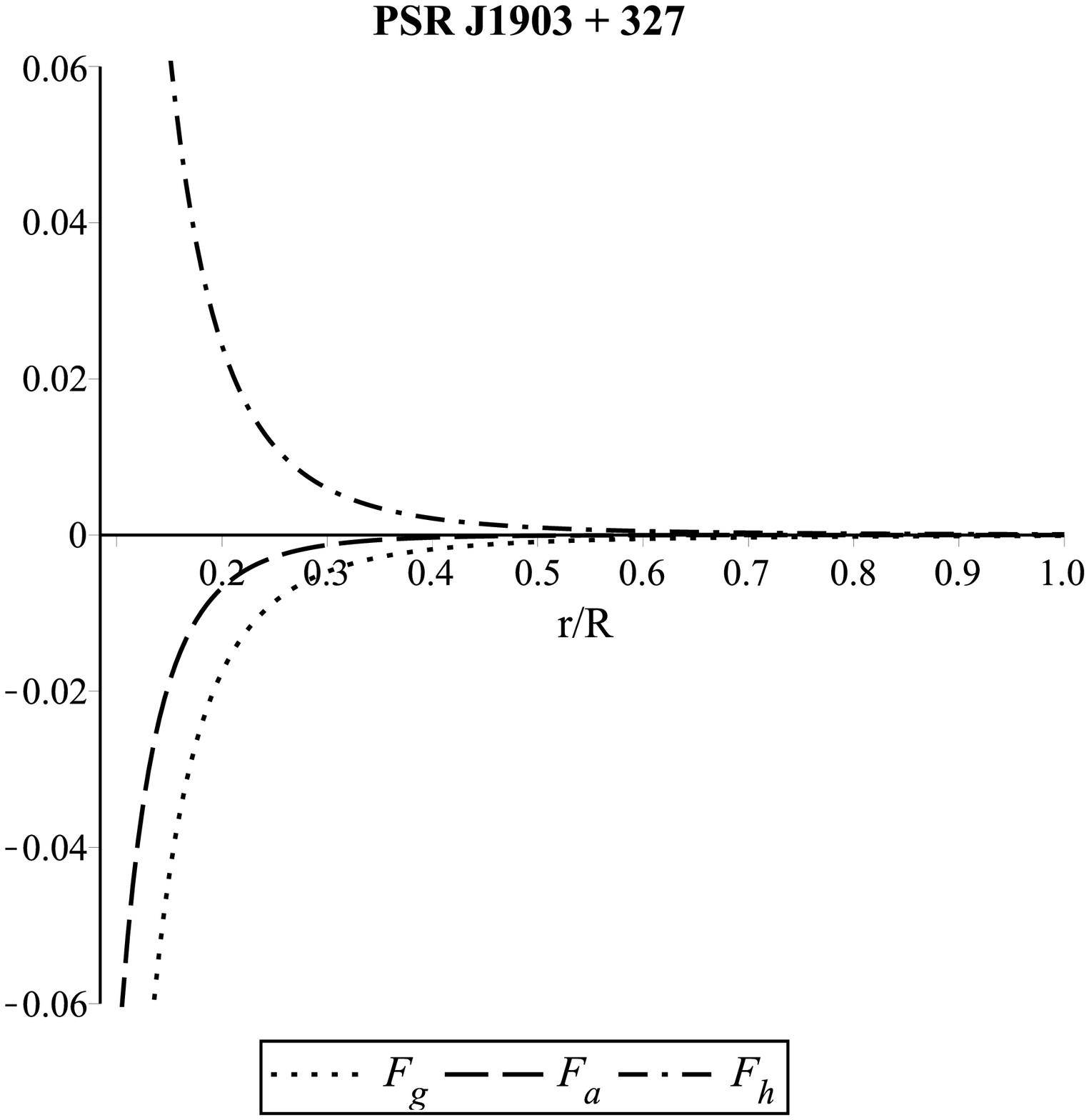}
\includegraphics[width=0.45\textwidth]{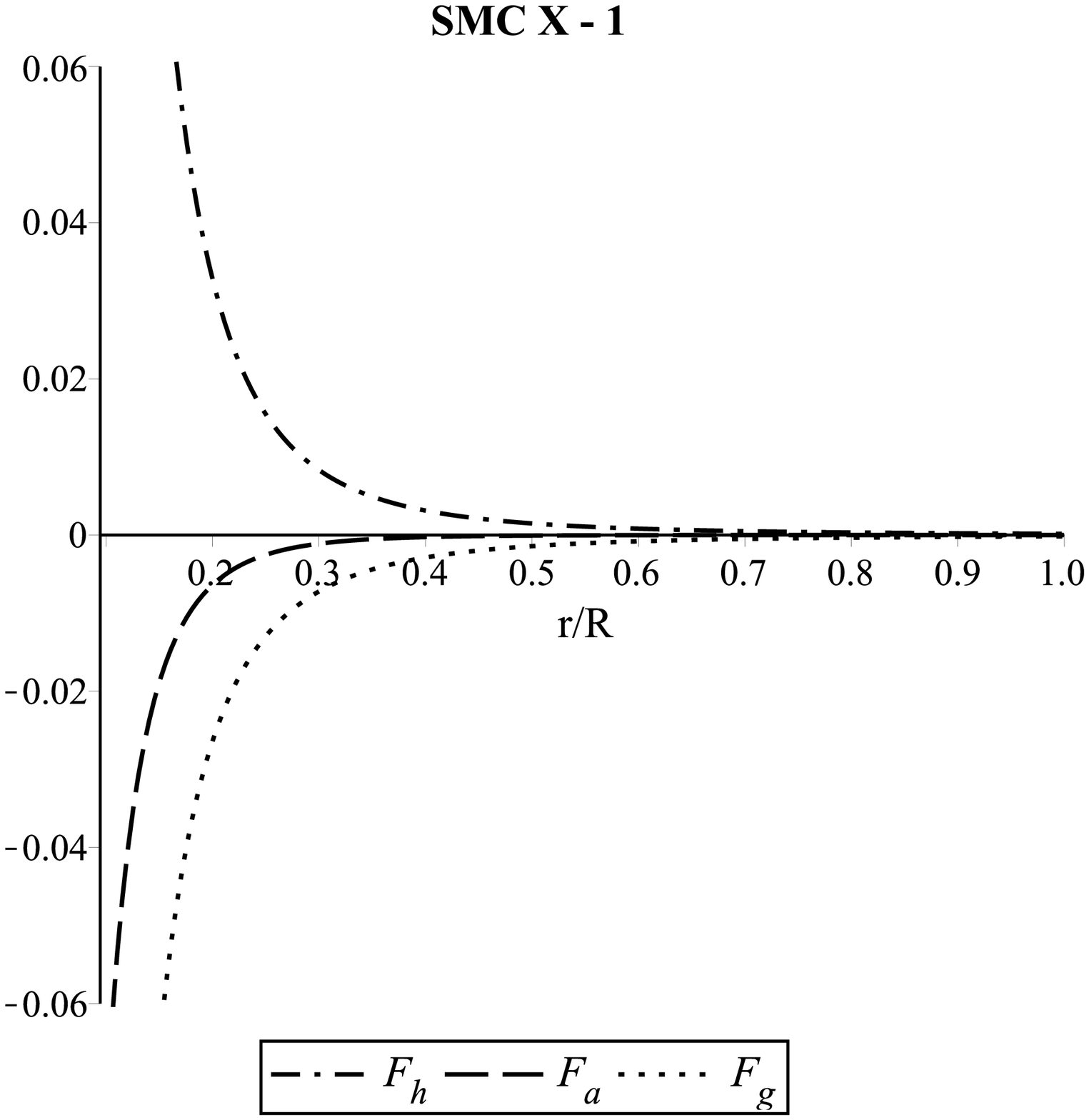}
\caption{Variation of the different forces with the fractional radial coordinate 
for different compact stars}\label{tov.}
\end{figure*}

Here $F_g$, $F_h$ and $F_a$ are gravitational, hydrostatics, and
anisotropic forces respectively. In Fig.~\ref{tov.} profile of the 
interaction between these three forces are shown in an elegent way. 
The figure indicates that the combined effects
of the gravitational and anisotropic forces is balanced by the hydrostatic
force which provides equilibrium configuration of the stellar structure.

\subsection{Herrera's condition for stability analysis}
According to causality condition the 
velocity of sound should follow the condition $0<v_s^{2}= dp/d\rho
<1$ for a physically realistic model as was proposed by \citet{Herrera1992} 
known as a technique for stability check of local anisotropic matter distribution. 
This technique dictates that the region for which radial speed of
sound is greater than the transverse speed of sound is a
potentially stable region. For our anisotropic
model, radial and transverse velocity of sound are defined by
\begin{eqnarray}
v_{rs}^{2}=\frac{dp_r}{d\rho}=\nonumber \\ \frac {1}{j(r)} \Bigg[\left\{ -3{r}^{4}f(r) n(r) +2
 \right\} \sqrt {1-rg(r) }\nonumber \\ +k \left\{ {r}^{4}f(r) n(r) +rg(r) -2 \right\} \Bigg],\label{40}
\end{eqnarray}

\begin{eqnarray}
v_{ts}^{2}=\frac{dp_t}{d\rho}=\nonumber \\ \frac {1}{j(r)} \Bigg[\left\{ -2{r}^{4} \left(h(r) -\frac{1}{2} \right) 
n(r) f(r) +{k}^{2}+1 \right\} \times  \nonumber \\ \sqrt {1-rg(r) }+2k \left\{ {r}^{4}f(r) n(r) +1/2\,rg(r) -1 \right\}\Bigg],\label{41}
\end{eqnarray}
where $n \left( r \right) ={\frac {b}{a{r}^{2}+1}}$, $f \left( r \right) ={\frac {a}{a{r}^{2}+1}}$, $h \left( r \right) ={\frac {a{r}^{2}+3}{a{r}^{2}+1}}$, $g \left( r \right) ={\frac {2\,br}{2\,a{r}^{2}+2}}$ and $j(r)={\sqrt {1
-rg \left( r \right) }{r}^{4}n \left( r \right) f \left( r \right) \left[ 2\,h \left( r \right) -1 \right] }$.

\begin{figure*}[thbp]
\centering
\includegraphics[width=0.4\textwidth]{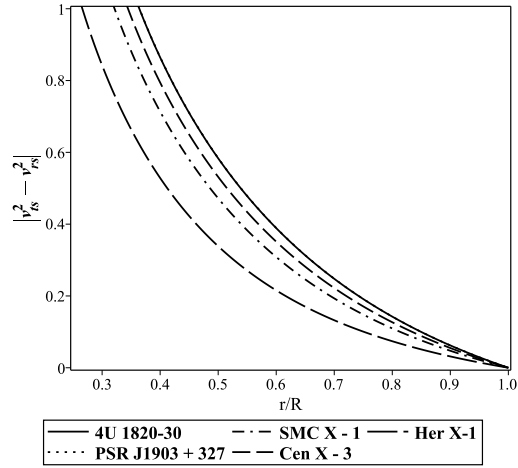}
\caption{Variation of the velocities, i.e. $v_{ts}^{2}-v_{rs}^{2}$ with the fractional radial coordinate for different compact stars}\label{vel.}
\end{figure*}

\begin{figure*}[thbp]
\centering
\includegraphics[width=0.4\textwidth]{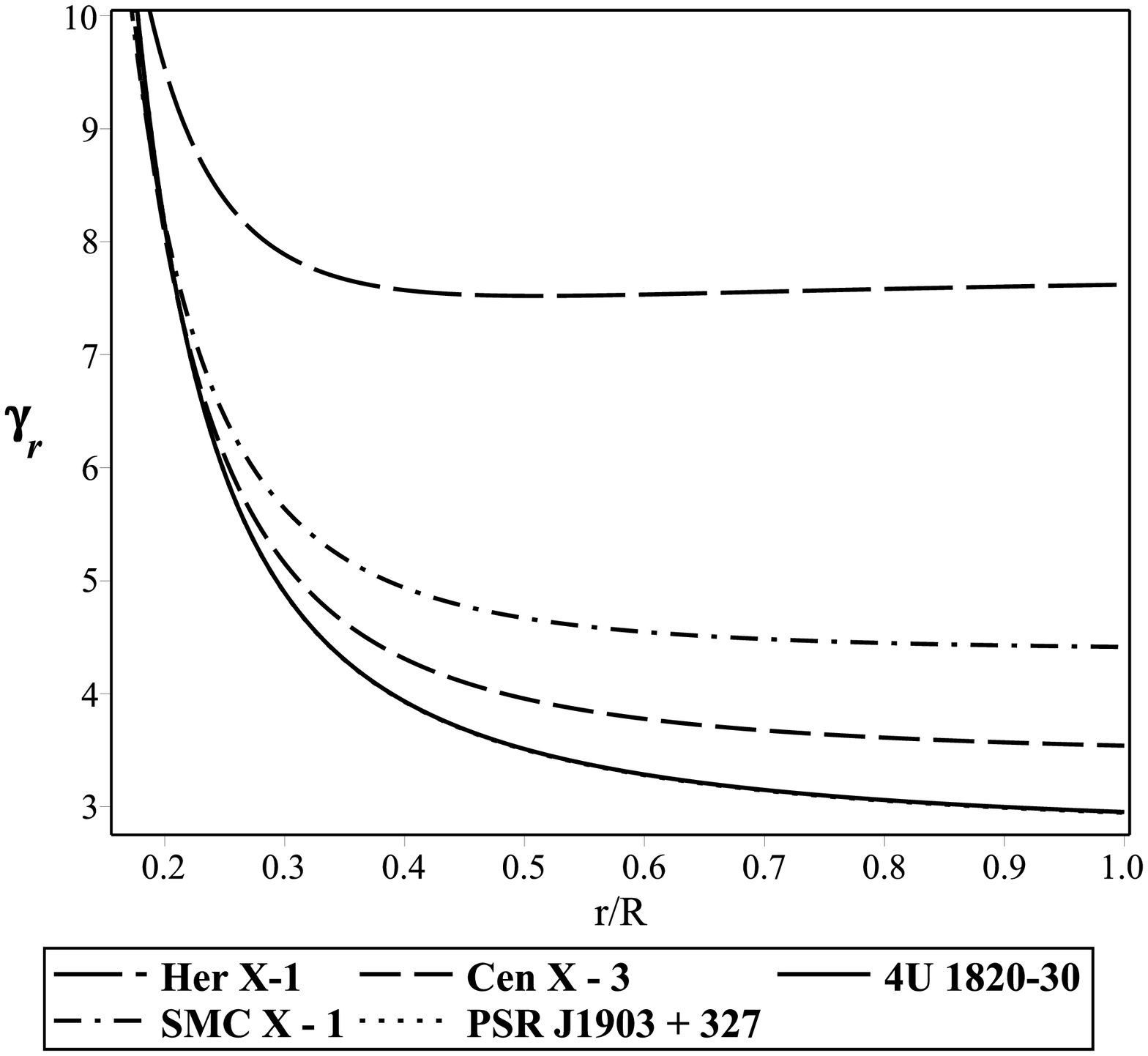}
\includegraphics[width=0.4\textwidth]{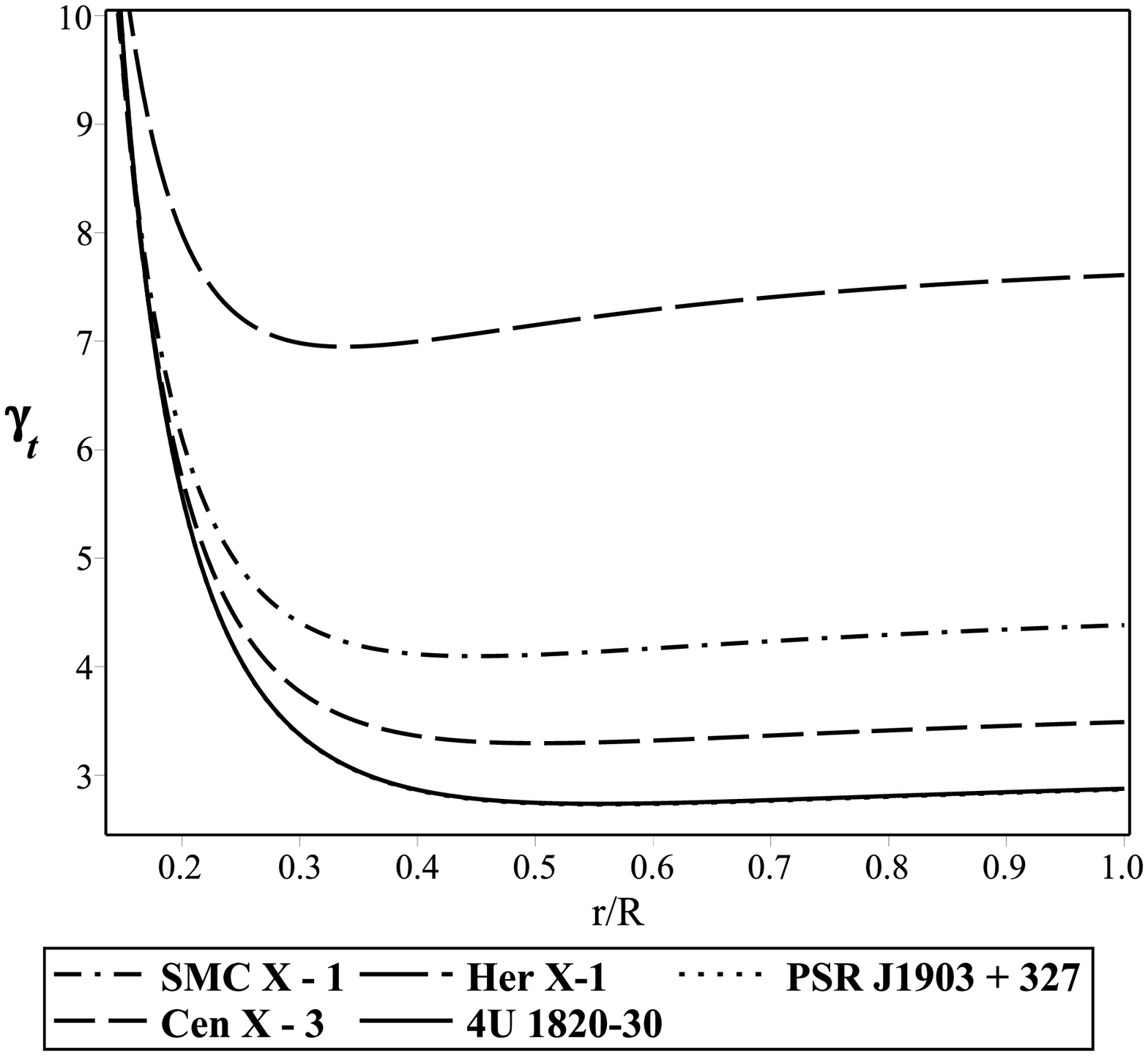}
\caption{Variation of the adiabatic index $\gamma_t$ and $\gamma_r$ 
with the fractional radial coordinate for $SMC~X-1$}\label{adia}
\end{figure*}

From the Fig.~\ref{vel.} it is very clear that for our model the radial and tangential speed
of sound, and also their difference consistent with the Herrera cracking concept but they do 
not satisfies causality condition. So in the aspect of causality condition we are not getting 
a stable stellar configuration. This is also true for the adiabatic index $\gamma_t$ and 
$\gamma_r$ as can be observed from Fig. 10.

\subsection{Compactification factor and redshift}
The mass of the compact star can be calculated from the density
profile
\begin{equation}
 m(r)=\frac{br^{3}}{2(1+ar^{2})}.\label{eqn42}
\end{equation}

\begin{figure*}[thbp]
\centering
\includegraphics[width=0.4\textwidth]{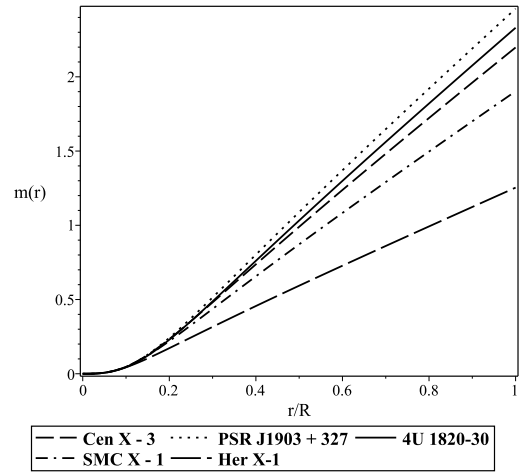}
\caption{Variation of the mass function with the fractional radial coordinate 
for different compact stars}\label{mass}
\end{figure*}

Now the mass function is regular at the origin as $r\rightarrow 0$
$m(r)\rightarrow 0$. The profile of the mass function shows (see Fig. 11) 
that mass function is monotonically increasing function of radius.

Since for a compact star, the maximum allowable ratio of the mass
to the radius cannot be arbitrarily large so by~\citet{Buchdahl1959} 
the ratio of twice the maximum allowable mass
to the radius is less than $8/9$, i.e. $2M/R<8/9$ where $M/R$ is
called the compactification factor which classifies the stellar
objects in different categories as shown by~\citet{Jotania2006}. 
It is to note that~\citet{Mak2001} derived a more simplified expression for the same ratio. 
In the present study we observe that all the stars satisfy the Buchdahl condition.

The compactification factor of our model is given by
\begin{equation}
u(r)=\frac{m(r)}{r}=\frac{br^{2}}{2(1+ar^{2})}.
\end{equation}\label{eqn43}

The variation of the compactification factor with the increasing
radius of the star is shown in Fig. 12 (left panel) from which it seems to be a
nonlinear increasing function of $r$, and get saturated after few
kilometer.

\begin{figure*}[thbp]
\centering
\includegraphics[width=0.4\textwidth]{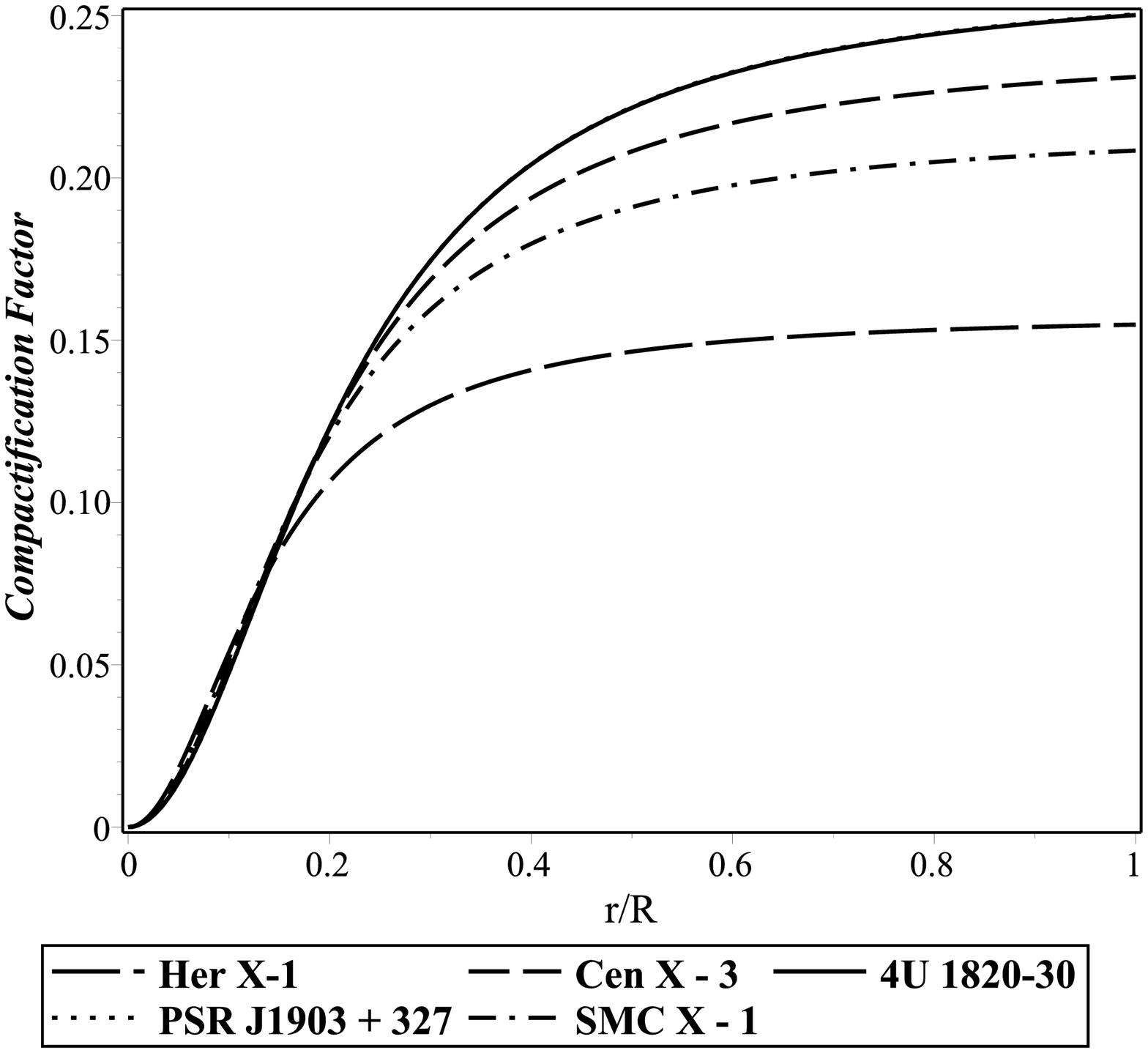}
\includegraphics[width=0.4\textwidth]{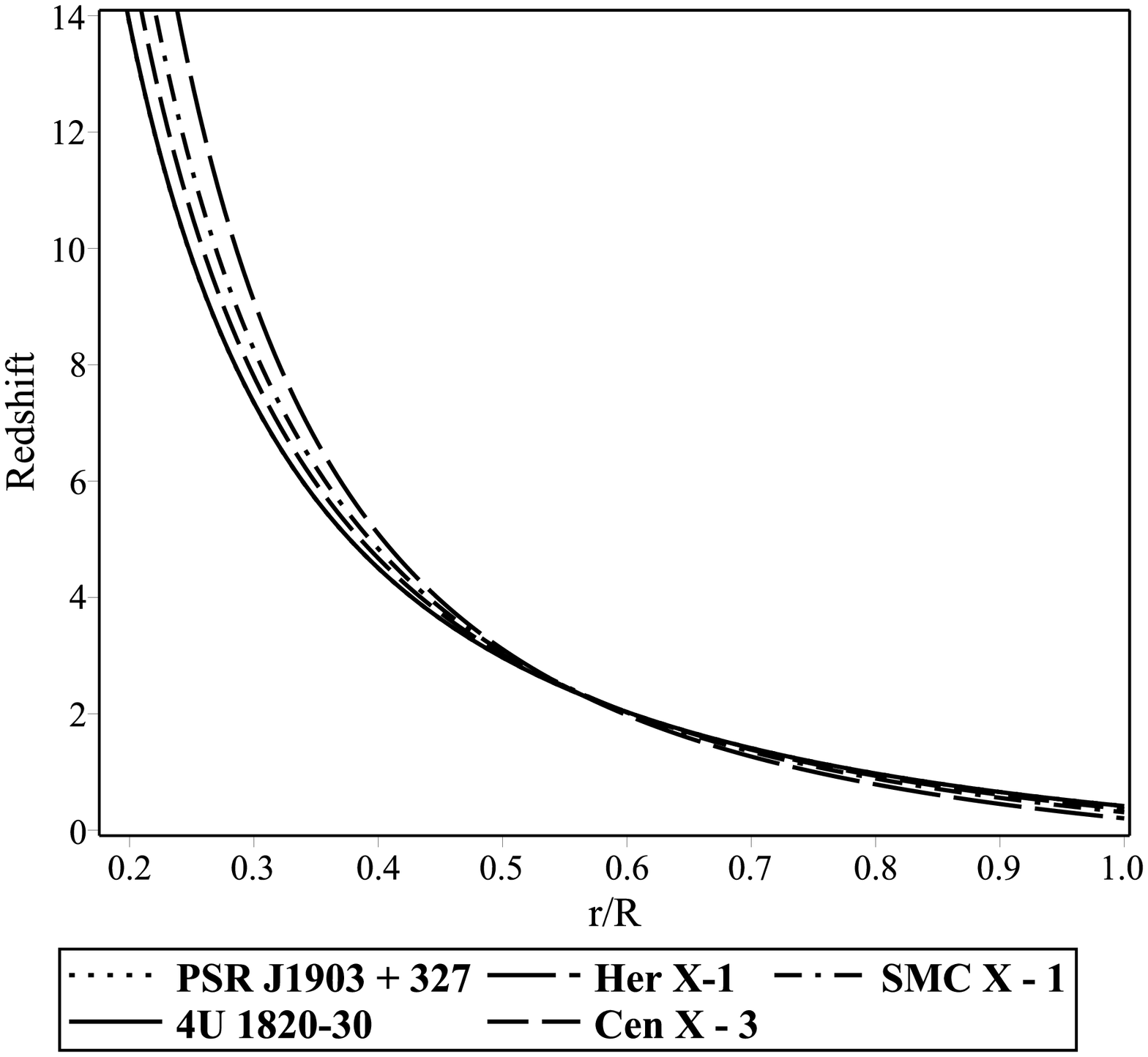}
\caption{Variation of the compactification factor (left panel) and the redshift (right panel) 
with the fractional radial coordinate for different compact stars}\label{comp.}
\end{figure*}

We also calculate the surface redshift of our model
\begin{equation}
Z_s=\left[{\frac {1+ \left( a-b \right)
{R}^{2}}{{R}^{2}a+1}}\right]^{-\frac{1}{2}}-1.
\end{equation}\label{eqn}

The surface redshift is plotted in Fig. 12 (right panel) and it is a decreasing
function of the radius of the star and it is also suffers from the
problem of singularity which is very off bit in nature.

\section{Concluding remarks}
In this present work we propose a new model of anisotropic compact
state with Matese and Whitman mass function under CKV. Though CKV 
formalism is suitable for solving non-linear differential equations, but the main
draw back of this tecnique is that it is not singularity free and hence the field 
equations contain contain singularity. Through this mathematical approach we have
highlighted different physical features of the compact anisotropic
object in terms of radial pressure, tangential pressure,
anisotropic factor, energy conditions of this spherical
distribution.

(i) The density profile, from the Matesae and Whitmann mass function,
have maximum density at the centre of the star and decreases
monotonically with the radius of the star (Fig. 1).

(ii) In our model the radial and transverse pressures suffer from the
problem of singularity which we suspect may be due to CKV
formalism. Both the pressures have maximum value at the centre and
decreases monotonically towards the surface (Figs. 2 and 3). It is noted that 
after $r=6.38$, the $p_t$ is greater than $p_r$, i.e. anisotropy is positive.

(iii) EOS parameters are shown in Fig. 4 which is satisfactory
in their nature.

(iv) The metric potentials increases non-linearly with the radius
of the star which is shown in Fig. 5 having finite value at $r=0$. The metric
functions $g_{rr}$, $g_{tt}$, $\frac{\partial {g_{tt}}}{\partial
r}$ are continuous at the boundary of the star. From these relations
we calculate the expressions as well as values of constants $a$,
$b$ and $k$.

(v) The anisotropic force is repulsive in nature which allows to
construct the more compact objects but does not vanishes at the
centre which is expected (Fig. 6).

(vi) All the energy conditions, i.e. SEC, NEC and
WEC are followed by our model as can be observed from Fig. 7. 
In this regard it is to note that our model satisfies SEC means 
the spacetime does contain a black hole region.

(vii) The matter distribution of the anisotropic star satisfies the
generalized TOV equation under gravitational force $F_g$, hydrostatic force
$F_h$ and anisotropic force $F_a$ (see Fig. 8). The combine effect of
the gravitational and anisotropic force is balanced by the hydrostatic
force. This result is in confirmation of the equilibrium of the system.

(viii) From the plot of difference of sound speeds it is clear that for 
the compact star $SMC~X-1$ according to Herrera's cracking concept 
we find stable region from 2.93 km to 9.13 km. In this connection Fig. 10 
is also notable one where both the radial and tangential adiabatic indices 
are greater than $\frac{4}{3}$ and hence confirms stability of the stellar model.

(ix) The mass function calculated here shows that it is regular
at the origin and also a monotonically increasing function of
radius (see Fig. 11). By usuing this mass function we have also shown 
the feature of compactification factor and redshift in Fig. 12. 

At the end of the article we have shown different physical properties, i.e., central~$({\rho}_c)$ and 
surface~$({\rho}_0)$ densities as well as surface redshift in Table \ref{Table 1}. 
From this table it is clear that our predicted model is consistent 
with the ultra dense compact and spherically symmetric stellar configurations.

\begin{table}
	\centering
	\caption{Physical Parameters for different compact stars from our model by using the data for mass and radius of different compact stars from~\citet{Rahaman2014} }\label{Table 1}
\begin{tabular}{ccccccccc}\hline 
Strange  stars &  Mass  & Radius& a & b & k & Surface Density & Central Density & Surface \\ 
   &  $\left({M}_{\odot}\right)$ & $\left(km\right)$ &  $\left({km}^{-2}\right)$  &  $\left({km}^{-2}\right)$&& $\left(gm/{{cm}^3}\right)$ & $\left(gm/{{cm}^3}\right)$  &  Redshift  \\
\hline 
$PSR~J1903+327$ & 1.667 & 9.82 & 0.23062 & 0.12068 & -0.35218 & $3.023\times {10}^{14}$ & $1.940\times {10}^{16}$ &           0.4152 \\ \hline 
$4U~1820-30$ & 1.58 & 9.316 &  0.25675 & 0.13422 & -0.35298 & $3.355\times {10}^{14}$ & $2.158\times {10}^{16}$ & 0.4146 \\\hline 
$Cen~X-3$ & 1.49 & 9.51 & 0.29018  & 0.13923 & -0.41824 & $2.940\times {10}^{14}$ & $2.239\times {10}^{16}$ & 0.3636 \\ \hline 
$SMC~X-1$ & 1.29 & 9.13 & 0.38270 & 0.16451 & -0.49078 & $2.843\times {10}^{14}$ & $2.645\times {10}^{16}$ &                            0.3094\\ \hline 
$Her~X-1$ & 0.85 & 8.1 & 0.79031 & 0.24937 & -0.64464 & $2.625\times {10}^{14}$ & $4.010\times {10}^{16}$ &                          0.2035\\ \hline 
      \end{tabular} 
\end{table}

\section*{Acknowledgments}
SR is thankful to the authority of the Inter-University Centre for
Astronomy and Astrophysics (IUCAA), Pune, India and Institute of
Mathematical Sciences (IMSc), Chennai, India for providing
Visiting Associateship under which a part of this work was carried
out.

\end{document}